\documentclass[prb,amsmath,amssymb,12pt]{revtex4}

\usepackage{graphicx}
\usepackage{dcolumn}
\usepackage{bm}

\begin{document}

\title{Interfacial instability induced by lateral vapor pressure fluctuation
in bounded thin liquid-vapor layers}

\author{Kentaro Kanatani}
\affiliation{%Division of Physics and Astronomy, Graduate School of Science,
Department of Physics,
Kyoto University, Kyoto 606-8502, Japan}

%\date{\today}

%\maketitle

\begin{abstract}
\thispagestyle{plain}
We study an instability of thin liquid-vapor layers 
bounded by rigid parallel walls from both below and above.
In this system, the interfacial instability is induced by
lateral vapor pressure fluctuation, 
which is in turn attributed to the effect of phase change:
evaporation occurs at a hotter portion of the interface
and condensation at a colder one.
The high vapor pressure pushes the interface downward and 
the low one pulls it upward.
A set of equations describing the temporal evolution of the interface 
of the liquid-vapor layers is derived.
This model neglects the effect of mass loss or gain at the interface
and guarantees the mass conservation of the liquid layer.
The result of linear stability analysis of the model shows that
the presence of the pressure dependence of the local saturation temperature
mitigates the growth of long-wave disturbances.
The thinner vapor layer enhances the vapor pressure effect.
We find the stability criterion, which suggests that
only slight temperature gradients are sufficient
to overcome the gravitational effect
for a water/vapor system.
%We also investigate the Rayleigh-Taylor instability of the system.
%The stabilizing vapor pressure effect is balanced
%with the destabilizing gravitational effect 
%in experimentally feasible systems.
The same holds for the Rayleigh-Taylor unstable case,
with a possibility that %the instability domain may be widened
the vapor pressure effect may be weakened
if the accommodation coefficient is below a certain critical value.
\end{abstract}

\maketitle\thispagestyle{plain}

\section{Introduction}

Thin liquid films have been intensively studied over the last decades.
Many contributions have been devoted to them, 
owing to their technological importance and wide industrial applications.
Their rich interfacial behaviors originate from combinations of various
effects such as capillarity, intermolecular forces, thermocapillarity
and gravity. 
One important effect among them is evaporation or condensation.
It was often incorporated in the studies of thin films,
\cite{burelbach88:_nonlin,oron97:_long}
and dewetting patterns resulting from drying of films were analyzed.
\cite{oron00:_three,schwartz01:_dewet,lyushnin02:_finger}

%In most works on evaporating or condensing films,
%the dynamics of the ambient gas was ignored
%and its pressure was assumed to be uniform,
%considering the situation of the infinitely deep gas phase.
%However, if the gas phase is confined and thin enough,
%the lateral pressure variation may occur
%and even affect the film dynamics.
Many past studies on evaporating or condensing liquid films
have ignored the dynamics of the gas above the liquid film,
assuming the infinitely deep gas phase.
\cite{burelbach88:_nonlin,oron97:_long,bestehorn06:_regul_rayleig_taylor}
However, in this study we consider a situation where the gas phase is
bounded by rigid parallel wall from above and has a finite depth
comparable with the liquid one.
%In this study, we address thin liquid-vapor layers 
%bounded by rigid parallel walls from both below and above.
%There are several papers on liquid-vapor layers.
The full linear stability analyses of this system
were performed in several papers.\cite{huang92:_instab,ozen04,ozen06:_rayleig_taylor,mcfadden07:_onset,mcfadden09:_onset_of_oscil_convec_in}
Despite the apparent simplicity of the configuration,
this system includes a free surface
and interfacial boundary conditions involving phase change,
and therefore is very sophisticated.
%so that only its linear stability was discussed until now.
In order to simplify this problem,
we apply long-wave approximation to both layers.
%at the cost of the neglect of convection.
%assuming that the thicknesses of the liquid and vapor layers are small
%enough to neglect convection.

The advantage of the use of long-wave or lubrication approximation
is the reduction of dimensionality:
a one-dimensional (two-dimensional) film evolution equation
can be derived in a two-dimensional (three-dimensional) system.
Normally, only the dynamics of the liquid is considered,
leading to a one-sided model.
However, if the ambient gas layer is thin enough,
%has a finite thickness,
a two-layer model would better describe the system.
This was demonstrated by VanHook et al.,\cite{vanhook97:_long_benar}
%who proposed an evolution equation of the interface of two fluid layers
who developed a two-layer theory
to reproduce their experimental results.
They showed that their two-layer model better predicts
the onset of instability in their experiment
than the corresponding one-layer model
and also correctly describes the formation of localized elevations.
In their approach, only the heat conduction in the gas phase
is taken into account, and the gas dynamics is ignored
because the viscosity of the gas is much less than that of the liquid.
Later, Merkt et al.\cite{merkt05:_long} 
presented an evolution equation
of the interface of two viscous fluid layers in the same geometry.
Their model allows for the shear stress 
induced by the motion of the upper layer
and therefore is reduced to the single layer equation
in the limit of small viscosity of the upper layer.
Although their goal is the observation of pattern formation
in the long-time regime, the two-layer models have been applied
to the cases of Rayleigh-Taylor instability\cite{burgess01:_suppr,alexeev07:_suppr_rayleig_taylor_maran}
and ultrathin films.\cite{joo00:_inter,lenz07:_compet}

\begin{figure}[htbp]
\begin{center}
 \includegraphics[scale=1.0]{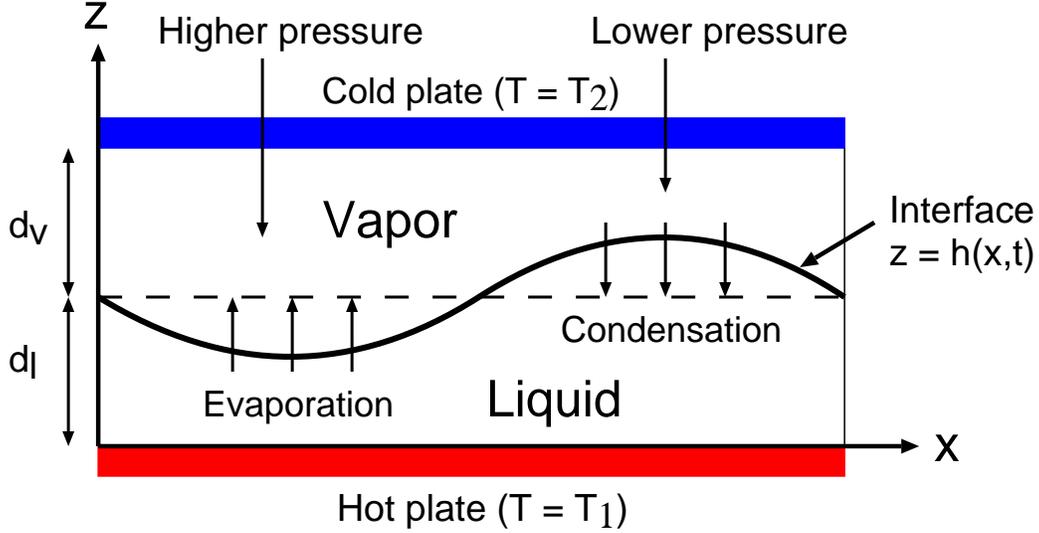}
\end{center}
 \caption{Instability mechanism of the present system.}
 \label{fig:system}
\end{figure}

Nevertheless, in the two-layer systems mentioned above
%do not undergo 
there is no phase transformation at the interface.
Here, we construct a two-layer theory for liquid-vapor layers
which undergo phase change, using long-wave approximation.
Note that the application of long-wave theory
to the vapor phase was made 
in the study of film boiling.\cite{panzarella00:_nonlin}
If we take into account the effect of the mass flux across the
interface, an instability peculiar to this system is expected,
even for the presence of large disparity in viscosity and density
between liquid and vapor; see Fig.~\ref{fig:system}.
The liquid film is initially in equilibrium with its vapor layer.
If the liquid side is heated or the vapor side is cooled,
evaporation occurs at a hotter portion of the interface
and condensation at a colder one.
Since the vapor layer is bounded,
the vapor pressure becomes higher in the evaporating region and
lower in the condensing one.
According to this lateral vapor pressure gradient,
the higher vapor pressure pushes the interface downward 
and the lower one pulls it upward.
Then, the surface deflection is amplified.
To our knowledge, this pressure-induced instability mechanism 
has not been considered in the past,
because the uniform ambient vapor pressure has been assumed
%in the previous works.
in previous studies of evaporating or condensing liquid films.
\cite{burelbach88:_nonlin,oron97:_long,bestehorn06:_regul_rayleig_taylor}

To derive the model, we require the interfacial boundary conditions
such that the mass transfer occurs between the phases.
We follow those of the earlier studies on evaporating or condensing liquid
films\cite{burelbach88:_nonlin,oron97:_long}
except the thermodynamic relation at the interface.
They used the linearized equation where the mass flux through the interface
is proportional to the difference between the interfacial temperature 
%deviation on the interface from 
and its saturation value corresponding to the surrounding vapor
pressure, based on kinetic theory.\cite{palmer76}
This relation cannot be directly applied to the present problem,
because the local saturation temperature varies
in the lateral direction,
depending on the vapor pressure. 
%in the lateral direction, as explained above.
Hence, we must modify the relation to take this effect into account.
Fortunately, this can be easily done in the thermodynamic framework.
For instance, Ajaev and Homsy\cite{ajaev01:_stead} and Wayner\cite{waynerCSA}
used a nonequilibrium thermodynamic relation including %the effect of 
the saturation temperature variation 
due to capillarity and disjoining pressure.
This effect was later included in the model
of evaporating or condensing thin liquid films.\cite{shklyaev07:_stabil}
However, they assumed the vapor pressure to be constant.
In our two-layer model, this relation should be extended
in accordance with the vapor pressure variation.
Thus, one of the purposes of this work is to investigate
the effect of the vapor pressure dependence of 
the local saturation temperature.
Note that this effect was not considered in the film boiling
case,\cite{panzarella00:_nonlin}
although the lateral vapor pressure variation, 
which drives the motion of the vapor,
certainly exists in the boiling film.
It is worthwhile noting that a somewhat similar motivation to ours
is found in Ref.~\onlinecite{sultan05:_evapor},
where the vapor concentration in the ambient gas phase 
above the liquid film fluctuates and thereby the mass flux varies 
along the interface.
However, they neglect the bulk gas dynamics itself
and consider only the diffusion of the vapor.

Here, we start with a more general nonequilibrium thermodynamic law,
which reduces to, in the linear domain, a proportional connection between 
the interfacial mass flux %is proportional to 
and the difference of chemical potential in each phase.\cite{colinet01:_nonlin_dynam_surfac_tension_driven_instab}
From this law, we can naturally derive a thermodynamic relation
similar to that of Ajaev and Homsy and Wayner. 
Moreover, the condition of local thermodynamic equilibrium, 
adopted in some papers on two-phase problem,\cite{ozen04,onuki05:_dropl,mcfadden07:_onset,mcfadden09:_onset_of_oscil_convec_in}
is recovered by taking the appropriate limit of the derived relation.
Therefore, the thermodynamic relation used here is also the extension
of the interfacial equilibrium condition into nonequilibrium states.
We note that more general formulation taking into account 
the nonequilibrium effect
contains a temperature discontinuity %between vapor and liquid 
at the liquid-vapor interface %under the presence of phase change,
during evaporation or condensation,
as was done in Ref.~\onlinecite{margerit03:_inter_benar_maran}.
However, this temperature jump %effect %emerges for rapid evaporation
%and hence it is neglected for our moderate evaporation case.
may be neglected unless the phase change occurs too rapidly.\cite{colinet01:_nonlin_dynam_surfac_tension_driven_instab}

In the derivation of our model,
we manipulate the mass flux balance equation at the liquid-vapor interface.
In the literature, the effect of mass loss or gain at the liquid surface
due to evaporation or condensation has been included
into the model through this equation by assuming that
the vapor speed is much larger than the liquid one
because the vapor density is much smaller than the liquid one.
\cite{burelbach88:_nonlin,oron97:_long}
However, in this paper we show another interpretation
of this equation as a consequence of order estimate.
%In the formulation, we assume that the vapor density, viscosity
%and thermal diffusivity are much smaller than those of the liquid.
%However, in order to take the vapor dynamics into account,
%we retain them in the long-wave scalings.
If we assume that the degree of the disparity in density 
between two phases is much greater than that in viscosity,
which is valid for most substances far from the critical point,
we can find that the liquid velocity is not balanced
by the vapor one in the mass balance equation.
Instead, it is balanced by the interface velocity,
which indicates that the effect of mass loss or gain can be neglected.
This order estimate leads to the approximation of the mass balance equation,
where the total mass of the liquid is conserved
and the effect of the mass flux affects only the dynamics of the vapor.
Under this approximation, we derive the model where the conservation
of the total liquid mass is guaranteed,
whereas the effect of evaporation or condensation
remains in the vapor dynamics.

The paper is organized as follows.
The model is formulated in Sec.~\ref{sec:formulation},
where the interfacial boundary condition
and scaling peculiar to this system mentioned above are introduced.
Linear stability results are presented in Sec.~\ref{sec:lsa},
including analyses of the Rayleigh-Taylor instability
and the effect of degree of nonequilibrium on it.
Section \ref{sec:conclusion} summarizes
the results and future work.

\section{\label{sec:formulation}Formulation}

For simplicity we consider a two-dimensional system 
as in Fig.~\ref{fig:system},
where the horizontal bilayers, liquid and vapor of the same substance, 
are confined by rigid parallel walls from both below and above.
We assume that the initial equilibrium thicknesses 
of the liquid and vapor layers,
$d_l$ and $d_v$, %respectively, 
are small enough to ignore the buoyancy effect.
The temperatures of the liquid-side and vapor-side plates are controlled
at $T_1$ and $T_2$. %, respectively.
The x axis is taken to be parallel to the walls,
and the z axis perpendicular to them.
The $z = 0$ plane corresponds to the boundary between the liquid
and the liquid-side plate.
The position of the liquid-vapor interface is described by $z = h(x,t)$.
Gravity acts in the negative direction of the z axis.

\subsection{Governing equations}

%\subsection{Vapor phase equations}
%
%\begin{eqnarray}
% \nabla \cdot \mathbf{v}_v &=& 0, \\
% \rho_v (\partial_t \mathbf{v}_v + \mathbf{v}_v \cdot \nabla \mathbf{v}_v)
% &=& - \nabla p_v + \eta_v \nabla^2 \mathbf{v}_v + \rho_v \mathbf{g}, \\
% \partial_t T_v + \mathbf{v}_v \cdot \nabla T_v &=& \kappa_v \nabla^2 T_v.
%\end{eqnarray}
%
%\subsection{Liquid phase equations}

We assume that the equations of continuity for incompressible fluids
and of the momentum and energy balance hold in each phase.
They are given by, respectively,
\begin{subequations}
\label{eq:governing_equations}
\begin{eqnarray}
 \nabla \cdot \mathbf{v}_{\beta} &=& 0, \label{eq:incompressible} \\
 \rho_{\beta} (\partial_t \mathbf{v}_{\beta} + \mathbf{v}_{\beta} \cdot \nabla \mathbf{v}_{\beta})
 &=& - \nabla p_{\beta} + \eta_{\beta} \nabla^2 \mathbf{v}_{\beta} -
 \rho_{\beta} g \mathbf{e}_z, \label{eq:NS} \\
 \partial_t T_{\beta} + \mathbf{v}_{\beta} \cdot \nabla T_{\beta} &=&
  \kappa_{\beta} \nabla^2 T_{\beta}. \label{eq:energy}
\end{eqnarray}
\end{subequations}
Here, $\mathbf{v}_{\beta} = (u_{\beta},w_{\beta})$, $p_{\beta}$ 
and $T_{\beta}$ are velocity, pressure and temperature fields,
respectively, in the $\beta = \{v,l\}$ phase,
where $v$ denotes the vapor and $l$ the liquid.
The differential operator is $\nabla \equiv (\partial_x,\partial_z)$.
%and $\Delta \equiv \nabla^2$.
The coefficients $\rho_{\beta}$, $\eta_{\beta}$ and $\kappa_{\beta}$ denote 
the density, dynamic viscosity and thermal diffusivity in the $\beta$ phase,
respectively, which are assumed to be constant in each phase.
In Eq.~(\ref{eq:NS}), $g$ is the gravitational acceleration 
and $\mathbf{e}_z$ the unit vector in the $z$ direction.

\subsection{Boundary conditions}

%\subsection{At the walls ($z = 0$ and $z = d_l + d_g$)}

At the walls ($z = 0$ and $z = d_l + d_g$),
we impose no-slip boundary conditions.
Along with the temperature conditions prescribed above, they read
\begin{subequations}
\label{eq:wall_boundary_conditions}
\begin{eqnarray}
 \mathbf{v}_l = \mathbf{0}, \quad %\\
 T_l = T_1 \quad &\textrm{at}& \quad z = 0, \label{eq:liquid_wall} \\
 \mathbf{v}_v = \mathbf{0}, \quad %\\
 T_v = T_2 \quad &\textrm{at}& \quad z = d_l + d_v. \label{eq:vapor_wall}
\end{eqnarray}
\end{subequations}
%\subsection{At the free surface ($z = h(x,y,t)$)}
At the liquid-vapor interface $z = h(x,t)$,
the mass flux $J$ must be conserved:
\begin{equation}
 J = \rho_v (\mathbf{v}_v \cdot \mathbf{n} 
  - \mathbf{v}_I \cdot \mathbf{n})
  = \rho_l (\mathbf{v}_l \cdot \mathbf{n} 
  - \mathbf{v}_I \cdot \mathbf{n}).
\label{eq:mass_flux}
\end{equation}
Here, $\mathbf{n}$ is the unit normal vector directed toward the vapor,
\begin{equation}
 \mathbf{n} = \frac{(-\partial_x h, 1)}
  {\sqrt{1 + (\partial_x h)^2}},
\end{equation}
and $\mathbf{v}_I$ represents the interface velocity,
which satisfies the kinematic condition
\begin{equation}
 \mathbf{v}_I \cdot \mathbf{n} = \frac{\partial_t h}
  {\sqrt{1 + (\partial_x h)^2}}.
\label{eq:kinematic}
\end{equation}
We assume the continuity of the tangential velocity along the interface,
\begin{equation}
 \mathbf{v}_v \cdot \mathbf{t} = \mathbf{v}_l \cdot \mathbf{t},
\label{eq:tangential_velocity}
\end{equation}
where $\mathbf{t}$ is the unit tangent vector to the interface,
\begin{equation}
 \mathbf{t} = \frac{(1, \partial_x h)}
  {\sqrt{1 + (\partial_x h)^2}}.
\end{equation}
%If evaporation and condensation are sufficiently slow,
%we may neglect any effects arising from 
%%molecular motion across the interface
%the discrepancy between the interface velocity and the bulk velocities
%at the interface and hence 
The interfacial stress and energy balance equations read,
%can be written as, 
respectively,\cite{burelbach88:_nonlin,oron97:_long}
\begin{eqnarray}
 &J (\mathbf{v}_l - \mathbf{v}_v) + (p_l - p_v) \mathbf{n} 
  - (2 \eta_l \mathbf{E}_l - 2 \eta_v \mathbf{E}_v) \cdot
  \mathbf{n} + 2 \sigma H \mathbf{n} = \mathbf{0},& \label{eq:stress} \\
 &J \{L + \frac{1}{2} [(\mathbf{v}_v - \mathbf{v}_I) \cdot
  \mathbf{n}]^2 - \frac{1}{2} [(\mathbf{v}_l - \mathbf{v}_I) \cdot
  \mathbf{n}]^2\} + \lambda_l \nabla T_l \cdot \mathbf{n} 
  - \lambda_v \nabla T_v \cdot \mathbf{n}& \nonumber \\
 &{} + [2 \eta_l \mathbf{E}_l \cdot (\mathbf{v}_l -
  \mathbf{v}_I) - 2 \eta_v \mathbf{E}_v \cdot (\mathbf{v}_v -
  \mathbf{v}_I)] \cdot \mathbf{n} = 0,& \label{eq:energy_balance}
\end{eqnarray}
where $\mathbf{E}_{\beta}$, $\sigma$, $H$, $L$ and $\lambda_{\beta}$
are the rate-of-strain tensor in the $\beta$ phase, the surface tension,
the mean curvature of the interface
\begin{equation}
 2 H = \frac{\partial_{x}^2 h}
  {[1 + (\partial_x h)^2]^{3/2}},
\end{equation}
the latent heat and the thermal conductivity in the $\beta$ phase,
respectively.
In Eq.~(\ref{eq:stress}), we ignore the thermocapillary (Marangoni)
term and assume $\sigma$ as well as $L$ and $\lambda_{\beta}$ 
to be constant for simplicity.
The Marangoni effect on two-phase surfaces has often been neglected in the
literature.\cite{huang92:_instab,panzarella00:_nonlin,lyushnin02:_finger,bestehorn06:_regul_rayleig_taylor} 
%as was recently justified for the linear problems
%in bilayer systems.
The recent investigations on the linearized systems of
liquid-vapor layers\cite{ozen04,mcfadden07:_onset,mcfadden09:_onset_of_oscil_convec_in}
and drop or bubble\cite{onuki05:_dropl},
show that the Marangoni effect
has little significance in pure two-phase coexisting states,
%for the ordinary material,
because the liquid-vapor interface becomes almost isothermal
owing to the large entropy difference between the two phases. 
In the Appendix, we include the thermocapillary term
in our model and examine its effect.
We numerically find that the thermocapillarity makes little contribution
to linear stability of a stationary state of the model
at least in the physical situations considered here.
%Nevertheless, we do not know whether the Marangoni effect is
%negligible in the nonlinear regime.
%Although we do not do this in the present paper,
%we can, in principle, add the Marangoni term 
%in Eq.~(\ref{eq:stress}) to examine its effect.
%Therefore, we .
The projection of Eq.~(\ref{eq:stress}) on the normal and tangent
to the interface yields, respectively,
\begin{subequations}
\label{eq:projections}
\begin{eqnarray}
 &J (\mathbf{v}_l - \mathbf{v}_v) \cdot \mathbf{n} + p_l - p_v 
  - \mathbf{n} \cdot (2 \eta_l \mathbf{E}_l - 2 \eta_v \mathbf{E}_v) \cdot
  \mathbf{n} + 2 \sigma H = 0,& \label{eq:normal_stress} \\
 &\mathbf{t} \cdot (2 \eta_l \mathbf{E}_l - 2 \eta_v \mathbf{E}_v) \cdot
  \mathbf{n} = 0,& \label{eq:tangential_stress}
\end{eqnarray}
\end{subequations}
where Eq.~(\ref{eq:tangential_velocity}) was used in the second equation.
Assuming the moderate phase change rate, 
the continuity of the temperature at the interface holds:
\begin{equation}
 T_l = T_v \equiv T_I.
\label{eq:temperature_continuity}
\end{equation}
Finally, in order to close the system we require an additional boundary
condition, which relates to the interfacial thermodynamic state.
In this study, we adopt the linearized phenomenological law
such that the mass flux across the interface is proportional to
deviation from local thermodynamic equilibrium:\cite{colinet01:_nonlin_dynam_surfac_tension_driven_instab}
\begin{equation}
 J = \hat{K} [\mu_l(p_l,T_I) - \mu_v(p_v,T_I)]. 
\label{eq:flux}
\end{equation}
Here, $\mu_{\beta}$ is the chemical potential in the $\beta$ phase,
which is a function of the pressure in the corresponding phase
and the temperature at the interface.
A proportionality coefficient $\hat{K}$ will be later specified 
by analogy with the kinetic theory.
We now expand the chemical potentials into Taylor series
%on the right hand side 
in this equation around their initial equilibrium value $\mu_0$ 
with respect to the variations of the pressure and the temperature,
\begin{subequations}
\begin{eqnarray}
 \delta p_{\beta} &=& p_{\beta} - p_0, \\
 \delta T &=& T_I - T_{sat}(p_0), \label{eq:dt}
\end{eqnarray}
\end{subequations}
where $T_{sat}(p_0)$ is the saturation temperature
at the initial equilibrium pressure $p_0$.
Using the Gibbs-Duhem relation for a one-component system, 
we obtain for each phase
\begin{eqnarray}
 \mu_{\beta}(p_{\beta},T_I) = %\mu_0 + \delta
%  \mu_{\beta}(p_{\beta},T_I) \\ \simeq
 \mu_0 - s_{\beta} \delta T + \frac{1}{\rho_{\beta}} \delta p_{\beta},
\label{eq:gdr}
\end{eqnarray}
where $s_{\beta}$ is the entropy density of the $\beta$ phase. %, and
%In Eq.~(\ref{eq:dt}), 
Substituting Eq.~(\ref{eq:gdr}) into Eq.~(\ref{eq:flux}) yields
\begin{equation}
 J = \hat{K} \left(\Delta s \delta T + \frac{1}{\rho_l} \delta p_l -
	     \frac{1}{\rho_v} \delta p_v\right),
\label{eq:clapeyron}
\end{equation}
where $\Delta s \equiv s_v - s_l$ is the entropy difference 
between the phases and related to the latent heat by
$L = T_{sat}(p_0) \Delta s$.
%\begin{equation}
% \Delta s = s_v - s_l.
%\end{equation}
%From Eq.~(\ref{eq:clapeyron}),
If we neglect the pressure terms on the right hand side 
of Eq.~(\ref{eq:clapeyron}), 
we recover the usual kinetic relation.\cite{palmer76,burelbach88:_nonlin,oron97:_long,panzarella00:_nonlin,bestehorn06:_regul_rayleig_taylor} 
%used in the investigations of thin films undergoing phase change at the interface
%which corresponds to the cases of the constant vapor pressure.
On the other hand, in the limit $\hat{K} \rightarrow \infty$,
Eq.~(\ref{eq:clapeyron}) reduces to the condition of local thermodynamic
equilibrium, used in several phase-boundary problems.\cite{ozen04,onuki05:_dropl,mcfadden07:_onset,mcfadden09:_onset_of_oscil_convec_in}
Therefore, Eq.~(\ref{eq:clapeyron}) is an intermediate relation
connecting the two different interfacial conditions appearing 
in the studies of two-phase systems with phase change.

\subsection{Dimensionless equations and parameters}

In order to nondimensionalize the above equations,
we scale lengths, time, velocities, pressures, temperatures and mass flux 
by $d_l$, $\displaystyle \frac{d_l^2 \rho_l}{\eta_l}$,
$\displaystyle \frac{\eta_l}{d_l \rho_l}$,
$\displaystyle \frac{\eta_l^2}{d_l^2 \rho_l}$, $\Delta T$ and
$\displaystyle \frac{\lambda_l \Delta T}{d_l L}$, respectively,
where $\Delta T$ is the initial temperature difference across the liquid layer.
We find $\Delta T$ together with $T_{sat}(p_0)$,
solving Eq.~(\ref{eq:energy}) for both phases with the boundary
conditions (\ref{eq:wall_boundary_conditions}), 
(\ref{eq:energy_balance}) and (\ref{eq:temperature_continuity})
in the equilibrium steady state ($J = 0$), as follows:
\begin{subequations}
\begin{eqnarray}
 &\Delta T \equiv T_1 - T_{sat}(p_0) = \displaystyle \frac{\lambda}{\lambda + d} 
  (T_1 - T_2),& \label{eq:temperature_difference} \\
 &T_{sat}(p_0) = \displaystyle \frac{\lambda T_2 + d T_1}{\lambda + d}.&
\end{eqnarray}
\end{subequations}
Here, the dimensionless parameters $\lambda$ and $d$ have been introduced.
The definitions of dimensionless parameters appearing in this paper
are presented in Table~\ref{tab:dimensionless_parameters}.
%and are defined in Table~\ref{tab:dimensionless_parameters}
%with those appearing below. 
Furthermore, we define the dimensionless pressure and temperature
such that their initial equilibrium values at the interface,
$p_0$ and $T_{sat}(p_0)$, correspond to $0$ in their new variables.
In the following, we show the resulting nondimensionalized equations.

%\subsection{Phase equations}

First, the governing equations of the liquid layer
(\ref{eq:governing_equations}) become
\begin{subequations}
\label{eq:nondimensional_liquid_governing_equations}
\begin{eqnarray}
 \nabla \cdot \mathbf{v}_l &=& 0, \label{eq:liquid_incompressible} \\
 \partial_t \mathbf{v}_l + \mathbf{v}_l \cdot \nabla \mathbf{v}_l
 &=& - \nabla p_l + \nabla^2 \mathbf{v}_l - G \mathbf{e}_z, 
 \label{eq:liquid_NS} \\
 P (\partial_t T_l + \mathbf{v}_l \cdot \nabla T_l) &=& \nabla^2 T_l, 
  \label{eq:liquid_energy}
\end{eqnarray}
\end{subequations}
and those of the vapor layer
\begin{subequations}
\label{eq:nondimensional_vapor_governing_equations}
\begin{eqnarray}
 \nabla \cdot \mathbf{v}_v &=& 0, \label{eq:vapor_incompressible}\\
 \rho (\partial_t \mathbf{v}_v + \mathbf{v}_v \cdot \nabla \mathbf{v}_v)
 &=& - \nabla p_v + \eta \nabla^2 \mathbf{v}_v - \rho G \mathbf{e}_z, 
 \label{eq:vapor_NS} \\
 P (\partial_t T_v + \mathbf{v}_v \cdot \nabla T_v) &=& \kappa \nabla^2 T_v.
  \label{eq:vapor_energy}
\end{eqnarray}
\end{subequations}

%\subsection{Boundary conditions}
The boundary conditions at the walls
(\ref{eq:wall_boundary_conditions}) reduce to
\begin{subequations}
\label{eq:nondimensional_wall_boundary_conditions}
\begin{eqnarray}
 u_l = w_l = 0, \quad
 T_l = 1 \quad &\textrm{at}& \quad z = 0,
 \label{eq:dimensionless_liquid_wall} \\
 u_v = w_v = 0, \quad
 \displaystyle T_v = -\frac{d}{\lambda} 
 \quad &\textrm{at}& \quad z = 1 + d, \label{eq:dimensionless_vapor_wall}
\end{eqnarray}
\end{subequations}
and those at the interface (\ref{eq:mass_flux}), 
(\ref{eq:kinematic}), (\ref{eq:tangential_velocity}),
(\ref{eq:projections}), (\ref{eq:energy_balance}), 
(\ref{eq:temperature_continuity})
and (\ref{eq:clapeyron}), respectively,
\begin{subequations}
\label{eq:nondimensional_interfacial_boundary_conditions}
\begin{eqnarray}
 &E J = \rho (\mathbf{v}_v \cdot \mathbf{n} 
  - \mathbf{v}_I \cdot \mathbf{n})
  = \mathbf{v}_l \cdot \mathbf{n} 
  - \mathbf{v}_I \cdot \mathbf{n},& \label{eq:dimensionless_mass_flux} \\
 &\mathbf{v}_I \cdot \mathbf{n} = \displaystyle \frac{\partial_t h}
  {\sqrt{1 + (\partial_x h)^2}},& \label{eq:dimensionless_kinematic} \\
 &\mathbf{v}_v \cdot \mathbf{t} = \mathbf{v}_l \cdot \mathbf{t},& \\
 &E J (\mathbf{v}_l - \mathbf{v}_v) \cdot \mathbf{n} + p_l - p_v 
  - \mathbf{n} \cdot (2 \mathbf{E}_l - 2 \eta \mathbf{E}_v) \cdot
  \mathbf{n} + 2 S H = 0,& \label{eq:dimensionless_normal_stress} \\
 &\mathbf{t} \cdot (2 \mathbf{E}_l - 2 \eta \mathbf{E}_v) \cdot
  \mathbf{n} = 0,& \label{eq:dimensionless_tangential_stress} \\
 &\displaystyle J + \frac{J \Pi \Theta}{2 E} \{[(\mathbf{v}_v 
  - \mathbf{v}_I) \cdot
  \mathbf{n}]^2 - [(\mathbf{v}_l - \mathbf{v}_I) \cdot
  \mathbf{n}]^2\} + \nabla T_l \cdot \mathbf{n} - \lambda \nabla T_v
  \cdot \mathbf{n}& \nonumber \\
 &\displaystyle {} + \frac{2 \Pi \Theta}{E^2} [\mathbf{E}_l \cdot (\mathbf{v}_l -
  \mathbf{v}_I) - \eta \mathbf{E}_v \cdot (\mathbf{v}_v -
  \mathbf{v}_I)] \cdot \mathbf{n} = 0,&
  \label{eq:dimensionless_energy_balance} \\
 &T_l = T_v = T_I,& \\
 &J = K \left[T_I + \displaystyle \frac{\Pi}{E} \left(p_l - \frac{1}{\rho}
			 p_v\right)\right].& \label{eq:dimensionless_clapeyron}
\end{eqnarray}
\end{subequations}
%Note that from Table~\ref{tab:dimensionless_parameters}
%the definition, %(\ref{eq:Phi}) 
%$\Pi$ does not depend
%on the temperature difference $\Delta T$. Then, 
%the temperature difference of the system $\Delta T$ is controlled
%only through the evaporation number $E$.
In Eq.~(\ref{eq:dimensionless_clapeyron}) we have introduced
the dimensionless parameter $K$,
%defined in Table~\ref{tab:dimensionless_parameters},
instead of $\hat{K}$ in Eq.~(\ref{eq:clapeyron}),
which has the dimension.
%whereas $K$ in Eq.~(\ref{eq:clapeyron}) has the dimension.
The parameters $K$ and $\hat{K}$ are related 
through $K = d_l L \hat{K} / \lambda_l$.
The value of the parameter $K$ 
defined in Table~\ref{tab:dimensionless_parameters}
is determined by the comparison with 
the kinetic theory (Hertz-Knudsen law).
%, which has been used in the literature.
%\cite{palmer76,burelbach88:_nonlin,oron97:_long,panzarella00:_nonlin,colinet01:_nonlin_dynam_surfac_tension_driven_instab,bestehorn06:_regul_rayleig_taylor}
In its definition, $\alpha$ is the accommodation coefficient,
$m$ is the molecular mass of the fluid and $k_B$ is the Boltzmann constant.
\begin{table}[htbp]
\caption{\label{tab:dimensionless_parameters}
Dimensionless parameters.
}
\begin{center}
\begin{tabular}{c@{\qquad}c@{\qquad}c@{\qquad}c}
\hline \hline \\[-5mm]
 Gravity & $\displaystyle G = \frac{g d_l^3 \rho_l^2}{\eta_l^2}$ &
 Density ratio & $\displaystyle \rho = \frac{\rho_v}{\rho_l}$ \\[5mm] 
 Liquid Prandtl number & $\displaystyle P = \frac{\eta_l}{\rho_l \kappa_l}$ &
 Dynamic viscosity ratio & $\displaystyle \eta = \frac{\eta_v}{\eta_l}$ \\[5mm] 
 Evaporation number & $\displaystyle E = \frac{\lambda_l \Delta T}{\eta_l L}$ & 
 Conductivity ratio & $\displaystyle \lambda = \frac{\lambda_v}{\lambda_l}$ \\[5mm]
 Surface tension & $\displaystyle S = \frac{\sigma \rho_l d_l}{\eta_l^2}$ & 
 Diffusivity ratio & $\displaystyle \kappa = \frac{\kappa_v}{\kappa_l}$ \\[5mm] 
 & $\displaystyle \Pi = \frac{\lambda_l \eta_l T_{sat}}{(L d_l \rho_l)^2}$ & 
 Initial thickness ratio & $\displaystyle d = \frac{d_v}{d_l}$ \\[5mm] 
 & $\displaystyle K = \alpha \frac{\rho_v d_l L^2}{\lambda_l T_{sat}}
 \sqrt{\frac{m}{2 \pi k_B T_{sat}}}$ & & 
 $\displaystyle \Theta = \frac{\Delta T}{T_{sat}}$ \\[5mm]
\hline \hline
\end{tabular}
\end{center}
\end{table}

\subsection{Long-wave asymptotics}

We apply the long-wave approximation to both layers.\cite{oron97:_long}
Letting a small parameter $\epsilon$ be $d_l / \Lambda$,
where $\Lambda$ represents the characteristic lateral length scale,
new space and time variables are introduced as
\begin{equation}
 x' = \epsilon x, \qquad z' = z, \qquad t' = \epsilon t.
\end{equation}
This rescaling indicates that the physical quantities vary much slower
in the horizontal direction than in the vertical one.
Assuming that $\epsilon \ll 1$, we expand the velocities, the pressures
and the mass flux in powers of $\epsilon$ as follows:
\begin{eqnarray}
\begin{array}{ll}
 u_l = u_{l_0} + \epsilon u_{l_1} + \ldots, &
 u_{v} = \epsilon^{-1} (u_{v_0} + \epsilon u_{v_1} + \ldots), \\
  w_l = \epsilon (w_{l_0} + \epsilon w_{l_1} + \ldots), &
  w_{v} = w_{v_0} + \epsilon w_{v_1} + \ldots, \\
 p_l = \epsilon^{-1} (p_{l_0} + \epsilon p_{l_1} + \ldots), &
 p_{v} = \epsilon^{-1} (p_{v_0} + \epsilon p_{v_1} + \ldots), \\
  J = J_{0} + \epsilon J_{1} + \ldots. &
\label{eq:expansion}
\end{array}
\end{eqnarray}
%Here, we assume that $u_l, J = \mathcal{O}(1)$.
%In consequence of this choice, $w_l = \mathcal{O}(\epsilon)$.
Here, we required $w_{\beta} / u_{\beta} = \mathcal{O}(\epsilon)$ %from
based on the continuity equations.
To take the pressure effects into account,
we chose $p_l, p_v = \mathcal{O}(\epsilon^{-1})$,
where the pressures of both layers are taken as the same order in $\epsilon$
because of the pressure balance.
From this and the balance between the pressure and viscous dissipation terms
in Eqs.~(\ref{eq:liquid_NS}) and
(\ref{eq:vapor_NS}), $|\mathbf{v}_l|$ and $\eta |\mathbf{v}_v|$
must be the same order.
Therefore, if we set $\eta = \mathcal{O}(\epsilon)$,
$|\mathbf{v}_l| / |\mathbf{v}_v| = \mathcal{O}(\epsilon)$ holds.
In order for all the physical effects to appear
in the leading-order equations,
some of the dimensionless parameters shown
in Table~\ref{tab:dimensionless_parameters}
are scaled by $\epsilon$ as
%Note that $u_v / u_l, w_v / w_l = \mathcal{O}(\epsilon^{-1})$,
%since we set $\eta = \mathcal{O}(\epsilon)$ in Eq.~(\ref{eq:vapor_NS}).
%For all the dimensionless parameters shown in
%Table~\ref{tab:dimensionless_parameters}, we adopt the scalings
\begin{eqnarray}
 \begin{array}{llllll}
  G = \epsilon^{-1} \tilde{G}, & P = \tilde{P}, & 
   E = \epsilon^2 \tilde{E}, & S = \epsilon^{-3} \tilde{S}, & 
   \Pi = \epsilon^5 \tilde{\Pi}, & K = \tilde{K}, \\
  \rho = \epsilon^2 \tilde{\rho}, & \eta = \epsilon \tilde{\eta}, &
   \lambda = \tilde{\lambda}, &
   \kappa = \epsilon^{-1} \tilde{\kappa}, & d = \tilde{d}, & \\
 \end{array}
\label{eq:scalings}
\end{eqnarray}
where we set $\Theta \rightarrow 0$ to
neglect the molecular kinetic energy and viscous dissipation terms
in Eq.~(\ref{eq:dimensionless_energy_balance}).
%and neglect the molecular kinetic energy and viscous dissipation terms
%in Eq.~(\ref{eq:dimensionless_energy_balance})
%by setting $\Theta \rightarrow 0$.
%to preserve the physical properties considered in this study.
The tildes denote the quantities of order $\mathcal{O}(1)$,
which will be used instead of the original dimensionless parameters.
Here we assumed that the evaporation number $E$ is of order
$\mathcal{O}(\epsilon^2)$, instead of $\mathcal{O}(\epsilon)$ as %was done 
in the work on evaporating or condensing liquid films.
\cite{burelbach88:_nonlin}
This is because in our system only a small amount
of evaporation or condensation is sufficient to drive the vapor dynamics
owing to the very small vapor density compared with the liquid density
(see a discussion below
Eqs.~(\ref{eq:leading_order_interfacial_boundary_conditions})
on decoupling of the mass balance equation
(\ref{eq:dimensionless_mass_flux})).
Since the density ratio $\rho$ is much smaller
than the dynamic viscosity one $\eta$
for most substances far from the critical point,
we set $\rho / \eta = \mathcal{O}(\epsilon)$.
Moreover, we took the thermal conductivity ratio $\lambda$
as $\mathcal{O}(1)$ in the scalings (\ref{eq:scalings}),
although the real value of $\lambda$ is very small
because in general the thermal conductivity of gas is much lower than
that of liquid.
We cannot take the limit $\lambda \rightarrow 0$
because the temperature boundary condition at the vapor-side wall
(\ref{eq:dimensionless_vapor_wall}) contains the factor $\lambda^{-1}$.
We substitute these scalings into the previous dimensionless equations
and take the limit $\epsilon \rightarrow 0$,
so that only the leading-order terms in $\epsilon$ are left in the
equations.
Hereafter, we shall omit the primes, the tildes and the subscripts $0$,
%wherever the confusion cannot occur.
%in the variables 
unless otherwise stated.

The leading-order governing equations are
\begin{subequations}
\label{eq:leading_order_liquid_equations}
\begin{eqnarray}
 &\partial_x u_{l} + \partial_z w_{l} = 0,& 
  \label{eq:leading_order_liquid_incompressible} \\
 &\partial_x p_{l} = \partial_z^2 u_{l},& 
  \label{eq:leading_order_liquid_NS_x} \\
 &\partial_z p_{l} + G = 0,& 
  \label{eq:leading_order_liquid_NS_z} \\
 &\partial_z^2 T_{l} = 0,& \label{eq:leading_order_liquid_energy}
\end{eqnarray}
\end{subequations}
for the liquid layer ($0 < z < h$)
from Eqs.~(\ref{eq:nondimensional_liquid_governing_equations}),
%(\ref{eq:liquid_incompressible})-(\ref{eq:liquid_energy}), 
and
\begin{subequations}
\label{eq:leading_order_vapor_equations}
\begin{eqnarray}
 &\partial_x u_v + \partial_z w_v = 0,& 
  \label{eq:leading_order_vapor_incompressible} \\
 &\partial_x p_v = \eta \partial_z^2 u_v,& 
  \label{eq:leading_order_vapor_NS_x} \\
 &\partial_z p_v = 0,& 
  \label{eq:leading_order_vapor_NS_z} \\
 &\partial_z^2 T_v = 0,& \label{eq:leading_order_vapor_energy}
\end{eqnarray}
\end{subequations}
for the vapor layer ($h < z < 1 + d$)
from Eqs.~(\ref{eq:nondimensional_vapor_governing_equations}). 
%(\ref{eq:vapor_incompressible})-(\ref{eq:vapor_energy}).
%Retaining $\epsilon$ associated with $\lambda$,
Whereas the wall boundary conditions 
(\ref{eq:nondimensional_wall_boundary_conditions})
%(\ref{eq:dimensionless_liquid_wall}) and (\ref{eq:dimensionless_vapor_wall}) 
remain unchanged,
those at the interface ($z = h$)
from Eqs.~(\ref{eq:nondimensional_interfacial_boundary_conditions})
%(\ref{eq:dimensionless_mass_flux})-(\ref{eq:dimensionless_clapeyron}) 
result in
\begin{subequations}
\label{eq:leading_order_interfacial_boundary_conditions}
\begin{eqnarray}
 &\partial_t h = - u_l \partial_x h + w_l,&
  \label{eq:liquid_mass_flux} \\
 &E J = \rho (- u_v \partial_x h + w_v),& \label{eq:vapor_mass_flux} \\
 &u_v = 0,& \label{eq:leading_order_tangential_velocity} \\
 &p_l - p_v + S \partial_x^2 h = 0,&
  \label{eq:leading_order_normal_stress} \\
 &\partial_z u_l = \eta \partial_z u_v&
  \label{eq:leading_order_tangential_stress} \\
 &J + \partial_z T_l - \lambda \partial_z T_v = 0,& 
  \label{eq:leading_order_energy_boundary} \\
 &T_l = T_v = T_I,& \label{eq:leading_order_temperature_continuity} \\
 &\displaystyle J = K \left(T_I - \frac{\Pi}{\rho E} p_v\right).&
  \label{eq:leading_order_clapeyron}
\end{eqnarray}
\end{subequations}
%where we have retained $\epsilon$ in the vapor temperature gradient term
%of Eq.~(\ref{eq:leading_order_energy_boundary}),
%because the temperature gradient in the vapor phase
%contains the $\mathcal{O}(1/\epsilon)$ term in the right expression
%of Eq.~(\ref{eq:temperature_gradient}) below.
%Both of the epsilons appearing 
%in Eqs.~(\ref{eq:leading_order_energy_boundary}) 
%and (\ref{eq:temperature_gradient}) originate from
%the rescaling of $\lambda$ (\ref{eq:scalings}) itself.
%In other words, if we assume $\lambda = \mathcal{O}(1)$,
%the epsilons in Eqs.~(\ref{eq:leading_order_energy_boundary}) 
%and (\ref{eq:temperature_gradient}) disappear.
%It gives the same result to take the limit
%$\lambda \rightarrow 0$ after %carrying out 
%the derivation with $\lambda = \mathcal{O}(1)$.
Note that from Eq.~(\ref{eq:dimensionless_mass_flux})
with Eq.~(\ref{eq:dimensionless_kinematic}) 
we obtain the two decoupled equations (\ref{eq:liquid_mass_flux}) 
and (\ref{eq:vapor_mass_flux}),
because the liquid and vapor velocities 
in Eq.~(\ref{eq:dimensionless_mass_flux}) 
do not balance under the scalings (\ref{eq:scalings}). 
%in $\epsilon$ assumed here.
%Since we have scaled the density ratio as $\mathcal{O}(\epsilon^2)$,
In other words, since we have scaled the density ratio 
as $\mathcal{O}(\epsilon^2)$,
the terms representing mass loss or gain become the next order 
in $\epsilon$ in Eq.~(\ref{eq:liquid_mass_flux}),
and hence are discarded there.
The next order mass balance gives Eq.~(\ref{eq:vapor_mass_flux}),
where the effect of the interface velocity
in Eq.~(\ref{eq:dimensionless_mass_flux})
has been neglected compared to that of the vapor velocity.
Therefore, Eq.~(\ref{eq:liquid_mass_flux})
implies that the total mass of the liquid layer is conserved 
at the leading order in $\epsilon$, 
while the effect of the mass flux affects only
the dynamics of the vapor layer through Eq.~(\ref{eq:vapor_mass_flux}).
%while the mass flux across the interface only affects 
%the dynamics of the vapor layer from Eq.~(\ref{eq:vapor_mass_flux}).
In addition, the vapor recoil term in the normal stress balance 
(\ref{eq:dimensionless_normal_stress}) and the liquid pressure term
in the thermodynamic relation (\ref{eq:dimensionless_clapeyron})
have disappeared with these scalings. 

The origin of the decoupling of the mass flux balance equation
(\ref{eq:dimensionless_mass_flux}) is more specifically explained as follows.
As is mentioned below Eq.~(\ref{eq:expansion}),
$|\mathbf{v}_l| / |\mathbf{v}_v|$ is the same order as
the dynamic viscosity ratio $\eta$.
%Since the density ratio $\rho$ is much less than $\eta$
%for most substances far from the critical point,
From the fact that $\rho \ll \eta$,
it follows that $\rho \ll |\mathbf{v}_l| / |\mathbf{v}_v|$
or $\rho |\mathbf{v}_v| \ll |\mathbf{v}_l|$.
For the second equality of Eq.~(\ref{eq:dimensionless_mass_flux}) to be true,
the liquid velocity $\mathbf{v}_l \cdot \mathbf{n}$ must be balanced
by the interface velocity
%This condition requires that $\mathbf{v}_l \cdot \mathbf{n}$
%in Eq.~(\ref{eq:dimensionless_mass_flux}) be equilibrated with
$\mathbf{v}_I \cdot \mathbf{n}$, which is represented
by Eq.~(\ref{eq:liquid_mass_flux})
through Eq.~(\ref{eq:dimensionless_kinematic}).
In Eq.~(\ref{eq:vapor_mass_flux}), $\mathbf{v}_I \cdot \mathbf{n}$
has disappeared because from Eq.~(\ref{eq:liquid_mass_flux})
it is the same order as $\mathbf{v}_l \cdot \mathbf{n}$,
much smaller than $\mathbf{v}_v \cdot \mathbf{n}$
%the interface velocity term has disappeared
%because the vapor velocity is much larger than the liquid velocity
from the small viscosity ratio $\eta \ll 1$.
Therefore, in order for the decoupling
of Eq.~(\ref{eq:dimensionless_mass_flux})
into Eqs.~(\ref{eq:liquid_mass_flux}) and (\ref{eq:vapor_mass_flux})
to be valid it is essential that $\rho \ll \eta \ll 1$ be the case.

Solving
Eqs.~(\ref{eq:leading_order_liquid_energy})
and (\ref{eq:leading_order_vapor_energy})
with the boundary conditions
(\ref{eq:nondimensional_wall_boundary_conditions})
%(\ref{eq:dimensionless_liquid_wall}), (\ref{eq:dimensionless_vapor_wall})
and
(\ref{eq:leading_order_temperature_continuity})
yields the temperature gradients in both layers
\begin{equation}
 \partial_z T_l = \frac{T_I - 1}{h}, \qquad
  \partial_z T_v = - \frac{T_I + \displaystyle d / \lambda}
  {1 + d - h}.
\label{eq:temperature_gradient}
\end{equation}
Substituting these equations into
Eq.~(\ref{eq:leading_order_energy_boundary})
and eliminating $J$ using Eq.~(\ref{eq:leading_order_clapeyron}),
we obtain
\begin{equation}
 K \left(T_I - \frac{\Pi}{\rho E} p_v\right) + \frac{T_I - 1}{h}
  + \frac{\lambda T_I + d}{1 + d - h} = 0.
\end{equation}
%Neglecting the retained $\mathcal{O}(\epsilon)$ term in this equation,
%Taking the limit $\lambda \rightarrow 0$,
Then, the surface temperature $T_I$ can be explicitly expressed as
\begin{equation}
 T_I = \frac{1}{\displaystyle 1 + K h + \frac{\lambda h}{1 + d - h}}
 \left[-\frac{(1 + d)(h - 1)}{1 + d - h}
   + \frac{K \Pi}{\rho E} h p_v\right].
\label{eq:interfacial_temperature}
\end{equation}
Substituting this equation into Eq.~(\ref{eq:leading_order_clapeyron})
again, we find the expression for the mass flux
\begin{equation}
 J = - \frac{K}{\displaystyle 1 + K h
  + \frac{\lambda h}{1 + d - h}}
  \left[\frac{(1 + d)(h - 1)}{1 + d - h} +
   \left(1 + \frac{\lambda h}{1 + d - h}\right)
   \frac{\Pi}{\rho E} p_v\right].
\label{eq:mass_flux_expression}
\end{equation}

%\begin{eqnarray}
% \begin{array}{ll}
%  \displaystyle u_l = \partial_x p_l \left(\frac{z^2}{2} - h z \right) 
%   + \frac{\eta}{2} \partial_x p_v (h - 1 - d) z, &
%   \displaystyle v_l = \partial_y p_l \left(\frac{z^2}{2} - h z \right) 
%   + \frac{\eta}{2} \partial_y p_v (h - 1 - d) z, \\[5mm]
%  \displaystyle u_v = \frac{1}{2} \partial_x p_v (z - 1 - d) (z - h), &
%   \displaystyle v_v = \frac{1}{2} \partial_y p_v (z - 1 - d) (z - h).
% \end{array}
%\end{eqnarray}

From Eqs.~(\ref{eq:leading_order_liquid_NS_z}) 
and (\ref{eq:leading_order_vapor_NS_z}),
we can find that the horizontal pressure gradients
$\partial_x p_l$ and $\partial_x p_v$ do not depend
on the vertical coordinate.
Then, we can twice integrate
Eqs.~%(\ref{eq:leading_order_liquid_incompressible}), 
(\ref{eq:leading_order_liquid_NS_x})
%(\ref{eq:leading_order_vapor_incompressible})
and (\ref{eq:leading_order_vapor_NS_x})
in the $z$ direction.
Using the boundary conditions
(\ref{eq:nondimensional_wall_boundary_conditions}),
%(\ref{eq:dimensionless_liquid_wall}), (\ref{eq:dimensionless_vapor_wall}),
(\ref{eq:leading_order_tangential_velocity}) and
(\ref{eq:leading_order_tangential_stress}), we obtain
\begin{eqnarray}
\begin{array}{l@{\hspace{9mm}}l}
 \displaystyle u_l = \frac{1}{2} \partial_x p_l z^2 + c_1 z, &
 \displaystyle \eta u_v = \frac{1}{2} \partial_x p_v (1 + d - z)^2 +
  c_2 (1 + d - z), %\\[3mm]
% \displaystyle w_l = - \frac{1}{6} \partial_x^2 p_l z^3 - \frac{1}{2} \partial_x c_1 z^2, &
%  \displaystyle \eta w_v = - \frac{1}{6} \partial_x^2 p_v (z - 1 - d)^3 - \frac{1}{2} \partial_x c_2 (z - 1 - d)^2,
\end{array}
\label{eq:explicit_horizontal_velocities}
\end{eqnarray}
with
\begin{equation}
  c_1(x,t) = -\displaystyle \frac{1}{2} (1 + d - h) \partial_x p_v -
  h \partial_x p_l, \qquad
  c_2(x,t) = \displaystyle \frac{1}{2} (1 + d - h) \partial_x p_v.  
\label{eq:coefficients}
\end{equation}
The expressions for the vertical velocities $w_l$ and $\eta w_v$
immediately follow from the integration of the continuity equations 
(\ref{eq:leading_order_liquid_incompressible})
and (\ref{eq:leading_order_vapor_incompressible})
with the no-slip boundary conditions
(\ref{eq:nondimensional_wall_boundary_conditions}):
%(\ref{eq:dimensionless_liquid_wall})
%and (\ref{eq:dimensionless_vapor_wall}):
\begin{eqnarray}
\begin{array}{l@{\hspace{9mm}}l}
% \displaystyle u_l = \frac{1}{2} \partial_x p_l z^2 + c_1 z, &
% \displaystyle \eta u_v = \frac{1}{2} \partial_x p_v (z - 1 - d)^2 + c_2 (z - 1 - d), \\[3mm]
 \displaystyle w_l = - \frac{1}{6} \partial_x^2 p_l z^3 - \frac{1}{2} \partial_x c_1 z^2, &
  \displaystyle \eta w_v = \frac{1}{6} \partial_x^2 p_v (1 + d - z)^3 -
  \frac{1}{2} \partial_x c_2 (1 + d - z)^2.
\end{array}
\label{eq:explicit_vertical_velocities}
\end{eqnarray}
%Regarding the pressure, integration of
%Eqs.~(\ref{eq:leading_order_liquid_NS_z})-(\ref{eq:leading_order_vapor_NS_z})
%with Eq.~(\ref{eq:leading_order_normal_stress}) gives
%the relation between the liquid and vapor pressure
%The pressures of both layers are related by 
Integration of
Eqs.~(\ref{eq:leading_order_liquid_NS_z}) and 
(\ref{eq:leading_order_vapor_NS_z})
with Eq.~(\ref{eq:leading_order_normal_stress}) gives
the relation between the liquid and vapor pressure
\begin{equation}
 \partial_x p_l = \partial_x p_v - S \partial^3_x h + G \partial_x h.
\label{eq:pressures_relation}
\end{equation}
%which is given by integration of
%Eqs.~(\ref{eq:leading_order_liquid_NS_z}) and 
%(\ref{eq:leading_order_vapor_NS_z})
%with Eq.~(\ref{eq:leading_order_normal_stress}).
%\begin{eqnarray}
% \partial_t h &=& - \partial_x \left(\frac{h^3}{6} \partial_x p_l +
%			      \frac{h^2}{2} c_1\right) \nonumber \\
% &=& \partial_x \left[\frac{h^3}{3} \partial_x p_l - \frac{h^2}{4}
%		 (h - 1 - d) \partial_x p_v\right] \\
% &=& \partial_x \left[\frac{h + 3 (1 + d)}{12} h^2 \partial_x p_v 
%		 + \frac{h^3}{3} \partial_x (G h - S \partial_x^2 h) \right] \nonumber
%% && + \partial_y \left\{\frac{h + 3 (1 + d)}{12} h^2 \partial_y p_v 
%%	     - \frac{S}{3} h^3 (\partial_x^2 +
%%	     \partial_y^2) \partial_y h \right\} \nonumber
%\end{eqnarray}
Substituting
Eqs.~(\ref{eq:explicit_horizontal_velocities}) 
and (\ref{eq:explicit_vertical_velocities})
with Eq.~(\ref{eq:coefficients})
into Eqs.~(\ref{eq:liquid_mass_flux}) and (\ref{eq:vapor_mass_flux})
%and eliminating $\partial_x p_l$ using Eq.~(\ref{eq:pressures_relation})
finally yields a set of equations, respectively,
\begin{eqnarray}
 \partial_t h &=& \partial_x \left[\frac{h + 3 (1 + d)}{12} h^2 \partial_x p_v 
		 + \frac{h^3}{3} \partial_x (G h - S \partial_x^2 h)
			   \right], \label{eq:film_thickness_equation} \\
% && + \partial_y \left\{\frac{h + 3 (1 + d)}{12} h^2 \partial_y p_v 
%	     - \frac{S}{3} h^3 (\partial_x^2 +
%	     \partial_y^2) \partial_y h \right\} \nonumber
 E J &=& - \frac{\rho}{12 \eta} \partial_x [(1 + d - h)^3 \partial_x p_v],
\label{eq:mass_flux_equation}
\end{eqnarray}
where the liquid pressure gradient $\partial_x p_l$ has been eliminated 
using Eq.~(\ref{eq:pressures_relation}).
%which has been simplified 
%making use of Eqs.~(\ref{eq:coefficients}) and (\ref{eq:pressures_relation}).
%\begin{eqnarray}
% \frac{\eta}{\rho} E J &=& - \partial_x \left[\frac{1}{6} (h - 1 - d)^3 
%		\partial_x p_v + \frac{1}{2} (h - 1 - d)^2 c_3\right]
% \nonumber \\
% &=& \frac{1}{12} \partial_x [(h - 1 - d)^3 \partial_x p_v]
%\end{eqnarray}
Equations (\ref{eq:mass_flux_expression}) and (\ref{eq:mass_flux_equation})
can be combined to eliminate $J$:
\begin{eqnarray}
 \hspace{-1cm} \frac{E(1 + d)(h - 1)}{1 + d - h}
  + \left(1 + \frac{\lambda h}{1 + d - h}\right) \frac{\Pi}{\rho} p_v
%  \hspace{10cm} \nonumber \\
 = \displaystyle \frac{\rho}{12 \eta}
 \frac{\displaystyle 1 + K h + \frac{\lambda h}{1 + d - h}}{K} 
 \partial_x [(1 + d - h)^3 \partial_x p_v]. 
\label{eq:vapor_pressure_equation}
\end{eqnarray}
%Here,
%\begin{equation}
%\tilde{\Phi} \equiv \frac{\tilde{E} \tilde{\Pi}}{\tilde{\rho}} = \epsilon^{-3} \frac{\lambda_l \eta_l T_{sat}}{\rho (L d_l \rho_l)^2},
%\label{eq:Phi}
%\end{equation}
%where the dropped tildes are restored to avoid the confusion.
%This new dimensionless parameter represents the variation 
%of the local saturation temperature due to the vapor pressure fluctuation.
Equations (\ref{eq:film_thickness_equation}) and
(\ref{eq:vapor_pressure_equation}) compose a closed system 
for the unknown variables $h(x,t)$ and $p_v(x,t)$.
The first term in square brackets of
Eq.~(\ref{eq:film_thickness_equation})
describes the effect of the lateral vapor pressure gradient,
while the second that of gravity and the surface tension.
The second term on the left hand side of
Eq.~(\ref{eq:vapor_pressure_equation})
represents that of the variation 
of the local saturation temperature due to the vapor pressure fluctuation.
Equation (\ref{eq:film_thickness_equation})
is written in the conserved form for $h$,
because we have neglected the effect of mass loss or gain
by decoupling the mass flux balance equation
(\ref{eq:dimensionless_mass_flux}) as before.
%which demonstrates that the mass conservation is guaranteed.
%It is worthwhile noting that this model cannot be reduced to any existing
%one-sided thin film equations with phase change\cite{burelbach88:_nonlin,bestehorn06:_regul_rayleig_taylor}
%by taking the limit $d \rightarrow \infty$,
%as can be done for the bounded two-layer models
%without phase change,\cite{vanhook97:_long_benar,merkt05:_long}
%owing to the effect of the lateral vapor pressure variation.
%because the mechanism of the film dynamics for our case 
%is essentially different from that for those single-layer ones.
%However, 
%If we set $\rho \rightarrow 0$,
%the vapor pressure vanishes according to
%Eq.~(\ref{eq:vapor_pressure_equation}),
%and hence the well-known thin film equation without phase change\cite{oron97:_long}
%is recovered from Eq.~(\ref{eq:film_thickness_equation}).
%since no mass loss or gain of the liquid layer occurs from Eq.~().
%This can be easily understood from Eqs.~(\ref{eq:liquid_mass_flux}) and (\ref{eq:vapor_mass_flux}).
%When $\rho \rightarrow 0$, $J \rightarrow 0$ from
%Eq.~(\ref{eq:vapor_mass_flux}),
%whereas no mass loss or gain of the liquid layer occurs from
%Eq.~(\ref{eq:liquid_mass_flux}).
%Note that we cannot take the limit $\eta \rightarrow 0$
%without taking the limit $\rho \rightarrow 0$ in our model,
%because we have assumed in Eq.~(\ref{eq:scalings}) 
%that the dynamic viscosity ratio is
%one-order lower in $\epsilon$ than the density ratio.

\section{\label{sec:lsa}Linear stability analysis}

The set of Eqs.~(\ref{eq:film_thickness_equation}) and
(\ref{eq:vapor_pressure_equation}) has a stationary solution
$h = 1$ and $p_v = 0$.
We perturb this state by
\begin{subequations}
\begin{eqnarray}
 h(x,t) &=& 1 + \hat{h} \exp(ikx + \omega t), \\
 p_v(x,t) &=& \hat{p} \exp(ikx + \omega t),
\end{eqnarray}
\end{subequations}
where $\hat{h}$ and $\hat{p}$ are infinitesimal quantities.
Linearizing the system gives the following growth rate: %dispersion relation:
\begin{equation}
 \omega = \frac{A k^2}{k^2 + k_0^2} - \frac{1}{3} k^2 (G + S k^2),
\label{eq:dispersion_relation}
\end{equation}
with
\begin{subequations}
\label{eq:Ak_0^2}
\begin{eqnarray}
 A &=& \frac{\eta (4 + 3d) (1 + d)}{\rho d^4}
 \frac{K}{\displaystyle 1 + K + \frac{\lambda}{d}} E, 
  \label{eq:A} \\
 k_0^2 &=& \frac{12 \eta (d + \lambda)}{\rho^2 d^4}
 \frac{K}{\displaystyle 1 + K + \frac{\lambda}{d}} \Pi. \label{eq:k_0^2}
\end{eqnarray}
\end{subequations}
Here, positive (negative) values of $\omega$ indicate instability growth
(decay).
The parameters $A$ and $k_0^2$ represent
the effect of lateral vapor pressure fluctuation
and that of local saturation temperature variation, respectively.
%due to the vapor pressure.
%In this subsection, we consider the thermodynamically unstable case ($A > 0$).
%The past dispersion relations of evaporating or condensing interfaces
%usually have the finite growth rate $\omega \neq 0$ at $k = 0$.\cite{burelbach88:_nonlin,panzarella00:_nonlin,bestehorn06:_regul_rayleig_taylor}
%The past film equation with phase change 
%entails the base state of uniformly evaporating boundary 
%or the finite growth rate $\omega > 0$ at $k = 0$
%when the liquid side is heated.\cite{burelbach88:_nonlin}
%However, in our case, 
From the dispersion relation (\ref{eq:dispersion_relation}),
one can easily find %see 
that the growth rate vanishes in the limit $k \rightarrow 0$.
Nevertheless, if $k_0^2$ were not present
in Eq.~(\ref{eq:dispersion_relation}),
the finite growth rate would remain for $k \rightarrow 0$.
%this would not be the case.
Therefore, the presence of the saturation temperature variation
mitigates long-wave growth rates.
Specifically, the saturation temperature is increased (decreased)
in higher (lower) vapor pressure regions
%which reduces the rate of evaporation or condensation.
and thereby the rate of evaporation or condensation is reduced.
This effect is prominent %in the long-wave regime.
for long-wave disturbances.
%if $k_0 \neq 0$.
%This is because no net mass exchange between the phases is considered
%in our model.
%there is no mass change for each phase,
%as mentioned above.
%This is a novel feature of our model,
%since 
%In contrast, 
%This is diferent from the other film equations with phase change,
%which have the nonzero growth rates at $k = 0$,
%\cite{burelbach88:_nonlin,panzarella00:_nonlin,bestehorn06:_regul_rayleig_taylor}
%%in the long-wave limit,
%as in the case of $k_0 = 0$ in Eq.~(\ref{eq:dispersion_relation}).
%Therefore, we can regard 
%The parameter $k_0^2$ can be regarded as the degree of the reduction
%of long-wave growth rates.
%From Eq.~(\ref{eq:k_0^2}) $k_0^2$ is proportional to $d_l / d_v^3$,
%because $\Pi$ is inversely proportional to $d_l^2$.
%This indicates that 
%Therefore, the reduction of long-wave disturbances
%is stronger for the thicker liquid layer and thinner vapor layer.
%This indicates that the presence of %the saturation temperature variation
%the pressure dependence of the local saturation temperature,
%denoted by $\Pi$,
%suppresses the growth of long-wave disturbances.
%This is a novel feature of our model,
%which has not been found in the previous two-phase problems 
%assuming the uniform ambient vapor pressure.
Notice that our model does not admit 
a quasisteady solution of flat moving interface
even if %the liquid side is heated or the vapor side is cooled.
$k_0 = 0$, because the lateral uniformity leads to $\partial_t h = 0$
from Eq.~(\ref{eq:film_thickness_equation});
this is a direct consequence of %the decoupling of
the neglect of mass loss or gain.
%Eq.~(\ref{eq:dimensionless_mass_flux})
%into Eqs.~(\ref{eq:liquid_mass_flux}) and (\ref{eq:vapor_mass_flux}).
We can consider only the long-wave limit $k \rightarrow 0$,
where $\omega \neq 0$ if $k_0 = 0$.

\subsection{Superheated or supercooled state}

\begin{figure}[htbp]
\scalebox{0.8}{
\includegraphics{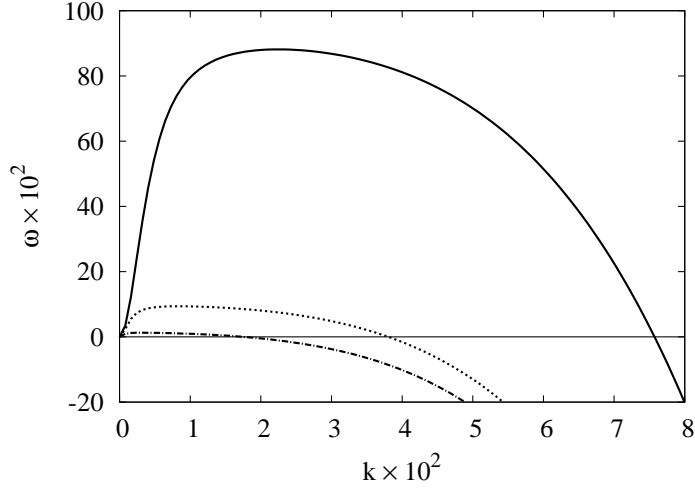}
}
\caption{\label{fig:dispersion}
Growth rates $\omega = \epsilon \tilde{\omega}$ versus wavenumber 
$k = \epsilon \tilde{k}$ for the water/vapor system 
at $100\ {}^\circ \mathrm{C}$ and 1 atm,
where $\tilde{\omega}$ and $\tilde{k}$ correspond to $\omega$ and $k$
in the dispersion relation (\ref{eq:dispersion_relation}).
Here $d_l = 10^{-4}$m, $\Delta T = 0.1\ {}^\circ \mathrm{C}$, %$G = 0$,
$g = 9.8\ \mathrm{m/s^2}$, $\alpha = 1$
%$K \rightarrow \infty$, $\epsilon = 10^{-2}$
and $d = 0.5$, $1$ and $2$ from top to bottom.
}
\end{figure}

To quantify the above results,
we consider the water/vapor system 
at $100\ {}^\circ \mathrm{C}$ and 1 atm.
Using the material properties shown in Table~\ref{tab:properties},
we plot the growth rates (\ref{eq:dispersion_relation}) 
in Fig.~\ref{fig:dispersion}, where we set $d_l = 10^{-4}$ m, 
$\Delta T = 0.1\ {}^\circ \mathrm{C}$,
$g = 9.8\ \mathrm{m/s^2}$ and $\alpha = 1$.
%If we take the accommodation coefficient $\alpha$ as unity,
%Since $K = 1.2 \times 10^3 \alpha$ for $d_l = 10^{-4}$m,
%we regard the coefficient $\displaystyle \frac{K}{1+K+\lambda/d}$ 
%in Eqs.~(\ref{eq:Ak_0^2}) as unity assuming $\alpha = 1$.
%$K \rightarrow \infty$ and $\epsilon = 10^{-2}$.
In the experiment, it is the temperature difference between the plates
$T_1 - T_2$, not across the liquid layer $\Delta T$
that can be controlled.
However, in the following we fix $\Delta T$
so that the vertical temperature gradient in each phase
is constant when we vary the value of the initial thickness ratio $d$.
%in order to maintain the vertical temperature gradients,
%which allows us to extract only the dependence 
%of the vapor layer thickness. %on the system.
From Eq.~(\ref{eq:temperature_difference})
$\Delta T = 0.1\ {}^\circ \mathrm{C}$ corresponds to
$T_1 - T_2 = 2.8\ {}^\circ \mathrm{C}$ for $d = 1$ in this system.
In the short-wave regime ($k \gg 1$),
the growth rate is negative because of the effect of the surface tension
($S$ in Eq.~(\ref{eq:dispersion_relation})),
whereas it is reduced in the long-wave regime ($k \ll 1$)
owing to that of the saturation temperature dependence on the vapor pressure
($k_0^2$ or $\Pi$ in Eqs.~(\ref{eq:dispersion_relation}) and
(\ref{eq:k_0^2})), as mentioned above.

\begin{table}[htbp]
\caption{\label{tab:properties}
Physical properties of water/vapor 
at $100\ {}^\circ \mathrm{C}$ and 1 atm,
identical with Table I of Ref.~\onlinecite{ozen04}.
}
\begin{center}
\begin{tabular}{l@{\hspace{1.5cm}}l@{\hspace{1.5cm}}l}
%\\[-5mm] 
\hline \hline \\[-7mm]
$\rho_l = 960 \mathrm{\ kg/m^3}$ & $\rho_v = 0.6 \mathrm{\ kg/m^3}$ &
$L = 2.3 \times 10^6 \mathrm{\ J/kg}$ \\[2mm]
$\eta_l = 2.9 \times 10^{-4} \mathrm{\ kg/m\ s}$ &
$\eta_v = 1.3 \times 10^{-5} \mathrm{\ kg/m\ s}$ &
$\sigma = 5.8 \times 10^{-2} \mathrm{\ N/m}$ \\[2mm]
$\lambda_l = 6.8 \times 10^{-1} \mathrm{\ J/m\ s\ {}^{\circ} C}$ &
$\lambda_v = 2.5 \times 10^{-2} \mathrm{\ J/m\ s\ {}^{\circ} C}$ & \\[2mm] 
$\kappa_l = 1.7 \times 10^{-7} \mathrm{\ m^2/s}$ &
$\kappa_v = 2.0 \times 10^{-5} \mathrm{\ m^2/s}$ &
% \\[2mm]
% & &
%$g = 9.8 \mathrm{\ m/s^2}$& 
 \\[4mm] \hline \hline
\end{tabular}
%\begin{tabular}{l@{\hspace{5mm}}l@{\hspace{5mm}}l@{\hspace{5mm}}l}
%\hline \hline \\
%$\rho_l = 960 \mathrm{\ kg/m^3}$ & $\rho_v = 0.6 \mathrm{\ kg/m^3}$ &
%$\eta_l = 2.9 \times 10^{-4} \mathrm{\ kg/m\ s}$ & 
%$\eta_v = 1.3 \times 10^{-5} \mathrm{\ kg/m\ s}$ \\[2mm]
%$\lambda_l = 6.8 \times 10^{-1} \mathrm{\ J/m\ s\ {}^{\circ} C}$ &
%$\lambda_v = 2.5 \times 10^{-2} \mathrm{\ J/m\ s\ {}^{\circ} C}$ & 
%$\kappa_l = 1.7 \times 10^{-7} \mathrm{\ m^2/s}$ &
%$\kappa_v = 2.0 \times 10^{-5} \mathrm{\ m^2/s}$ \\[2mm]
%$L = 2.3 \times 10^6 \mathrm{\ J/kg}$ & 
%$\sigma = 5.8 \times 10^{-2} \mathrm{\ N/m}$ & &
%%$g = 9.8 \mathrm{\ m/s^2}$& 
% \\[4mm] \hline \hline
%\end{tabular}
\end{center}
\end{table}

The three dispersion curves with different values of $d$ 
in Fig.~\ref{fig:dispersion} suggest that
the instability is enhanced for the thinner vapor layer,
which reflects the fact that
$A$ in the dispersion relation (\ref{eq:dispersion_relation})
is a monotonically decreasing function of $d$ according to Eq.~(\ref{eq:A}).
As can be seen from Eq.~(\ref{eq:dispersion_relation}),
the factor $A$ roughly represents the intensity of the instability
unless $k_0^2$ is very large.
Physically, the friction at the wall of the vapor side prevents the vapor flow
from mitigating the lateral vapor pressure gradient,
and this effect is more pronounced for the narrower vapor layer.
Therefore, the destabilizing effect of the lateral vapor pressure
variation is stronger for the thinner vapor layer.
The stabilizing role of the vapor flow is also understandable by considering
the viscosity of the vapor.
From Eqs.~(\ref{eq:Ak_0^2}),
$A$ and also $k_0^2$ are proportional to the viscosity ratio $\eta$,
which implies that increasing the vapor viscosity
intensifies the instability.
The large viscosity of the vapor weakens the vapor flow
from the equation of the vapor motion
(\ref{eq:leading_order_vapor_NS_x})
and also from the tangential stress boundary condition
(\ref{eq:leading_order_tangential_stress}).
Thus, the reduction of the destabilizing vapor pressure effect by the
vapor flow is suppressed.
Although both decreasing $d$ and increasing $\eta$
impede the vapor flow, the dependence of $A$ or $k_0^2$ on them
is different, because the vapor flow is prevented by the different mechanism.

The above argument is confirmed by directly estimating the vapor flow.
Integration of the horizontal component of the vapor velocity
(\ref{eq:explicit_horizontal_velocities}) in the vertical direction
gives the expression for the total vapor flux
\begin{equation}
  \label{eq:vapor_flow}
  q_v = \int^{1 + d}_h u_v dz
  = -\frac{1}{12 \eta} (1 + d - h)^3 \partial_x p_v.
\end{equation}
Similarly, the liquid flux is obtained as
\begin{equation}
  \label{eq:liquid_flow}
  q_l = \int^h_0 u_l dz
  = -\frac{h + 3(1 + d)}{12} h^2 \partial_x p_v
  - \frac{1}{3} h^3 \partial_x (G h - S \partial_x^2 h).
\end{equation}
In terms of $q_v$ and $q_l$, the model equations
(\ref{eq:film_thickness_equation}) and (\ref{eq:mass_flux_equation})
are rewritten as
\begin{eqnarray}
  \label{eq:rewritten}
  \partial_t h = -\partial_x q_l, \qquad \qquad
  \frac{E J}{\rho} = \partial_x q_v.
\end{eqnarray}
If we regard $h$ as nearly $1$,
the vapor and liquid fluxes are
\begin{equation}
  \label{eq:flows}
  q_v \simeq -\frac{d^3}{12 \eta} \partial_x p_v, \qquad \qquad
  q_l \simeq -\frac{4 + 3 d}{12} \partial_x p_v
  - \frac{1}{3} \partial_x (G h - S \partial_x^2 h).
\end{equation}
Substituting Eqs.~(\ref{eq:flows}) into Eqs.~(\ref{eq:rewritten}),
one can find the origin of the intensity of the instability $A$:
the factor $\eta / d^3$ in the expression for $A$ (\ref{eq:A})
arises from the vapor flux and $4 + 3 d$ from the liquid flux.
The rest is a contribution from phase change or the temperature difference.
Thus, it is shown that
the stabilizing vapor flow is responsible for 
the strong dependence of the instability intensity
on the initial thickness ratio
and also for that on the viscosity ratio.
%is attributed to 
%the roles of the vapor and liquid flows in the instability
%are shown.
%We note that the instability is also strengthened by decreasing
%the liquid viscosity,
%because it increases the mobility of the liquid.

It is worthwhile noting that
the role of the vapor flow presented here
is different from that described %of Ozen and Narayanan
in the study of full linear stability analysis
of a similar bilayer system, where the penetration of the fluid
at the walls is allowed. \cite{ozen04}
%including the effect of convection.
%where the stabilizing effect of the vapor flow
%and the destabilizing effect of the liquid flow
%is attributed to the convection of heat.
The authors of Ref.~\onlinecite{ozen04} 
attributed the stabilizing effect of the vapor flow
to the convection of heat: the vapor flow
%explained that the vapor flow stabilizes the instability because it
convects heat from a hotter portion of the interface
to a colder one.
Their description cannot be applied to our system,
because the term representing heat convection does not appear
in the leading-order equations, (\ref{eq:leading_order_liquid_energy})
and (\ref{eq:leading_order_vapor_energy}),
of the long-wave approximation.
%since heat conduction prevails against heat convection
%under the long-wave scaling.
%since we ignore the effect of heat convection.
%and therefore the lateral vapor pressure variation should have a critical role 
%in causing the instability.

%\begin{eqnarray}
% \omega = \frac{E (4 + 3d)(1 + d) k^2}
%  {\displaystyle \frac{12 d \Phi}{\rho} +
%  \frac{\rho}{\eta} \frac{1 + K}{K} d^4 k^2} - \frac{S}{3} k^4
%\end{eqnarray}
From Eq.~(\ref{eq:dispersion_relation}),
we can determine the cutoff wavenumber $k_c$ analytically.
By setting $\omega = 0$, we obtain
\begin{equation}
 k_c^2 = \left(\frac{k_0^2}{2} + b\right)
  \left[- 1 + \sqrt{1 + 2 \frac{a - b k_0^2}
   {(k_0^2 / 2 + b)^2}}\ \right],
\label{eq:cut-off}
\end{equation}
where
\begin{equation}
 a = \frac{3 A}{2 S}, \qquad b = \frac{G}{2 S}.
\end{equation}
If the right hand side of Eq.~(\ref{eq:cut-off})
does not take a real positive value,
%does not give the real solution for $k_c$,
the cutoff wavenumber $k_c$ does not exist
and hence the dispersion curve never crosses the line $\omega = 0$.
This indicates that the system is linearly stable,
because from Eq.~(\ref{eq:dispersion_relation}) the surface tension
makes the growth rate negative in the short-wave limit 
$k \rightarrow \infty$.
Since $k_c^2$ is negative when $a - b k_0^2 < 0$,
the criterion for the linear stability is expressed as %equivalent to
\begin{equation}
 \frac{b k_0^2}{a} = \frac{G k_0^2}{3 A}
 = \frac{4 (d + \lambda)}{(4 + 3 d)(1 + d)} \frac{G \Pi}{\rho E} > 1.
\label{eq:criterion}
\end{equation}
%which represents the competition of the two terms in the numerator
%inside the root of Eq.~(\ref{eq:cut-off}).
This inequality is equivalent to the condition that
the growth rate is negative at infinitesimal wavenumber,
which can be found by Taylor expansion of Eq.~(\ref{eq:dispersion_relation})
around $k = 0$.
This stability condition %(\ref{eq:criterion})
is independent of the factors representing the vapor flux
contribution to $A$, $\eta / d^3$, %mentioned above,
and the effect of degree of nonequilibrium,
$\displaystyle \frac{K}{1 + K + \lambda / d}$,
because $k_0^2$ also has the same factors and they are canceled out. 
%This stability criterion 
%is independent of the surface tension $S$ and
%the degree of nonequilibrium $K$.
%Assuming 
For the same condition as before, 
%$d_l = 10^{-4}$ m, %$\Delta T = 0.01\ {}^\circ \mathrm{C}$,
%$d = 1$ and $g = 9.8\ \mathrm{m/s^2}$ in the system considered, % here,
the left hand side of Eq.~(\ref{eq:criterion}) becomes %equals
$7.3 \times 10^{-4}$ %, while the right hand side
%while the evaporation number $1.0 \times 10^{-5} \epsilon^{-2}$ 
for $d = 1$. % and $\Delta T = 0.01\ {}^\circ \mathrm{C}$.
%$2.2 \times 10^{-3}\ \epsilon^{-2}$.
%These values are far from the satisfaction of the criterion 
%for the linear stability 
%(\ref{eq:criterion}).
%Therefore, %it is concluded that 
%gravity has virtually no influence
%on the stability of the realistic system. %ordinary case.
From this evaluation, it is concluded that only slight temperature gradients
are sufficient to overcome the stabilizing gravitational effect
in the realistic system.
Nevertheless, it is noted that the unstable modes
of infinitesimal wavenumbers can be eliminated
by appropriately modulating the horizontal scale of the system.
Therefore, the stability criterion for such a system 
should be weaker than Eq.~(\ref{eq:criterion}).
%which implies that the influence of phase change on the instability 
%is negligible.
%\begin{equation}
% k_c^2 = \frac{G + S k_0^2}{2 S}
%  \left\{- 1 + \sqrt{1 - 4 S \frac{G k_0^2 - 3 A}
%   {(G + S k_0^2)^2}}\right\}
%\end{equation}
%\begin{eqnarray}
% k_{max}^2 &=& \left[\frac{1}{6}\left\{\frac{2 k_0^6 - 6 b k_0^4 
%				 + 3 (2 b^2 + 9 a) k_0^2 - 2 b^3}{9} 
%				 + \sqrt{\frac{a k_0^2 (4 k_0^6 
%				 - 12 b k_0^4 + 3 (4 b^2 + 9 a) k_0^2 
%				 - 4 b^3}{3}}\right\}\right]^{\frac{1}{3}}
% \nonumber \\
% & & {} + \left[\frac{1}{6}\left\{\frac{2 k_0^6 - 6 b k_0^4 
%				 + 3 (2 b^2 + 9 a) k_0^2 - 2 b^3}{9} 
%				 - \sqrt{\frac{a k_0^2 (4 k_0^6 
%				 - 12 b k_0^4 + 3 (4 b^2 + 9 a) k_0^2 
%				 - 4
%				 b^3}{3}}\right\}\right]^{\frac{1}{3}}
% \nonumber \\
% & & {} - \frac{2 k_0^2 + b}{3}
%\end{eqnarray}
%\begin{eqnarray}
% \lefteqn{k_{max}^2 = - \frac{2 k_0^2 + b}{3}} \nonumber \\
% {} &+& \left\{\frac{2 k_0^6 - 6 b k_0^4 
%	 + 3 (2 b^2 + 9 a) k_0^2 - 2 b^3}{54} 
%	 + \frac{1}{6} \sqrt{\frac{a k_0^2 [4 k_0^6 
%	 - 12 b k_0^4 + 3 (4 b^2 + 9 a) k_0^2 
%	 - 4 b^3]}{3}}\right\}^{1 / 3}
% \\
% {} &+& \left\{\frac{2 k_0^6 - 6 b k_0^4 
%	 + 3 (2 b^2 + 9 a) k_0^2 - 2 b^3}{54} 
%	 - \frac{1}{6} \sqrt{\frac{a k_0^2 [4 k_0^6 
%	 - 12 b k_0^4 + 3 (4 b^2 + 9 a) k_0^2 
%	 - 4 b^3]}{3}}\right\}^{1 / 3}
% \nonumber
%\end{eqnarray}

The expression for the fastest growing mode $k_{max}$ 
can be also analytically obtained
from Eq.~(\ref{eq:dispersion_relation}).
Straightforward calculation yields
\begin{eqnarray}
 k_{max}^2 = - \frac{2 k_0^2 + b}{3} %\nonumber \\
 &+& \left\{\frac{(k_0^2 - b)^3}{27} + \frac{1}{2} a k_0^2 
    + \frac{1}{6} \sqrt{\frac{a k_0^2 [4 (k_0^2 - b)^3 + 27 a
    k_0^2]}{3}}\right\}^{1 / 3}
 \nonumber \\
 &+& \left\{\frac{(k_0^2 - b)^3}{27} + \frac{1}{2} a k_0^2 
    - \frac{1}{6} \sqrt{\frac{a k_0^2 [4 (k_0^2 - b)^3 + 27 a
    k_0^2]}{3}}\right\}^{1 / 3}.
\label{eq:fastest}
\end{eqnarray}
Unlike the results of linear stability analysis of the other film equations,
a simple relation between $k_c$ and $k_{max}$ cannot be established
for our model.
Equation (\ref{eq:fastest}) seems to be too complicated
to find any asymptotic form.
However, in the special case $k_0^2 = b$,
it reduces to a simple form
\begin{equation}
 k_{max}^2 = - b + (a b)^{1 / 3}.
\label{eq:simple_fastest}
\end{equation}
This special case is realizable %for our system
for the water/vapor system considered %realistic value 
if we set $d_l = 2.3 \times 10^{-5}$ m for $d = 1$ and $K \rightarrow \infty$.
%From Eqs.~(\ref{eq:cut-off}) and (\ref{eq:fastest}), 
%and (\ref{eq:simple_fastest}),
%gravity turns out to be an important factor
%in determining the cutoff and fastest growing wavenumbers, 
%although we have illustrated that it almost does not change 
%the occurrence of the instability itself. % of the system.
%\begin{equation}
% \omega = \frac{A k^2}{k^2 + k_0^{2*} / S^2} 
%  - \frac{S}{3} k^2 (G^* S^2 + k^2)
%\end{equation}

%\subsection{Gravity-free case}

%\subsection{Effect of gravity}

%\begin{equation}
% k_c^2 = \frac{G^* S^4 + k_0^{2*}}{2 S^2}
%  \left\{- 1 \pm \sqrt{1 - 4 S^3 \frac{G^* k_0^{2*} S - 3 A}
%   {(G^* S^4 + k_0^{2*})^2}}\right\}
%\end{equation}

%\begin{equation}
% G^* k_0^{2*} S < 3 A
%\end{equation}

\subsection{Rayleigh-Taylor instability}

We also investigate the Rayleigh-Taylor instability of the system,
\cite{bestehorn06:_regul_rayleig_taylor,panzarella00:_nonlin,merkt05:_long,alexeev07:_suppr_rayleig_taylor_maran,mcfadden07:_onset,burgess01:_suppr,ozen06:_rayleig_taylor}
where gravity acts toward the vapor side.
%by changing the sign of gravity.
%where the liquid layer overlies the vapor one.
The system of interest is the case that
the layers are heated from below or cooled from above,
so that the stabilizing effect of evaporation or condensation
counteracts the destabilizing one of gravity.
In this subsection, we consider the balance between these two effects
by changing the signs of gravity and the temperature difference.

Before starting the analysis, we show the difference from 
the model of evaporating or condensing liquid films 
with infinitely deep vapor layer.
\cite{bestehorn06:_regul_rayleig_taylor}
In Ref.~\onlinecite{bestehorn06:_regul_rayleig_taylor},
the authors made the two assumptions:
the much larger gas layer depth than the liquid one
and the neglect of the latent heat in the temperature boundary condition
at the interface (\ref{eq:energy_balance})
(Eqs.~(3) or (4) in Ref.~\onlinecite{bestehorn06:_regul_rayleig_taylor}).
To compare their dispersion relation with ours,
we abandon these assumptions and show their dispersion relation
derived without imposing them.
%The growth rate of that model has the same form 
%as Eq.~(\ref{eq:dispersion_relation}) with $A,G < 0$ given that $k_0 = 0$.
%However, the definition of $A$ is different.
In our notation, it reads
\begin{equation}
  \label{eq:dispersion_BM}
   \omega_{BM} = - |A_{BM}| + \frac{1}{3} k^2 (G - S k^2),
\end{equation}
with
%$A$ in Ref.~\onlinecite{bestehorn06:_regul_rayleig_taylor} corresponds to
\begin{equation}
 |A_{BM}| = \frac{1 + d}{d}
 \frac{K}{\displaystyle 1 + K + \frac{\lambda}{d}} |E|.
%\epsilon^2 \tilde{K} |\tilde{E}| = \mathcal{O}(\epsilon^2),
\label{eq:A_BM}
\end{equation}
The assumptions which they made are equivalent to $d \gg 1$ and $K \ll 1$.
Hence, by taking these limit we obtain $ |A_{BM}| = K |E|$,
which is consistent with the result of
Ref.~\onlinecite{bestehorn06:_regul_rayleig_taylor}.
However, for the water/vapor system considered here
$K = 1.2 \times 10^3 \alpha$ for $d_l = 10^{-4}$m.
Therefore their assumption
of the negligible latent heat ($K \ll 1$) is questionable
unless the accommodation coefficient $\alpha$ is very small.
Their dispersion relation (\ref{eq:dispersion_BM}) has the same form
as ours (\ref{eq:dispersion_relation}), 
except for the presence of $k_0^2$.
%given that $k_0 = 0$.
%where the tildes are restored.
However, their definition of $A$ in Eq.~(\ref{eq:A_BM}) is different
from ours (\ref{eq:A})
%This difference stems from
because of the different mechanism:
local mass loss or gain at the interface is the main stability mechanism
in Ref.~\onlinecite{bestehorn06:_regul_rayleig_taylor}.
In contrast, this effect has been neglected in our system
compared to that of the vapor pressure fluctuation.
This is also confirmed by taking the ratio between both $A$:
$\displaystyle \frac{|A_{BM}|}{|A|} = \frac{\rho}{\eta} \frac{d^3}{4 + 3 d}$.
Note that this ratio is independent of $K$.
For the water/vapor system, $|A_{BM}| / |A| \simeq 0.011$
%this ratio becomes nearly equal to unity
for $d = 2$.
Therefore $|A_{BM}| \ll |A|$ holds in the system considered here.
%as long as
%the liquid and vapor layer thicknesses are comparable and
Here the condition for the neglect of
the effect of mass loss or gain, $\rho \ll \eta$, is required,
as was mentioned before.
%emerged again.
However, if $d$ is much larger,
the effect of mass loss or gain can be comparable to
that of vapor pressure fluctuation,
because the latter effect is much more weakened
for a thicker vapor layer.
%and comparable to that of mass loss or gain.
%the neglect is valid for arbitrary values of $K$.
%in our system is reasonable.
%as can also be seen from the fact that Eq.~(\ref{eq:BM}) is 
%of order $\mathcal{O}(\epsilon^2)$.
%in the first approximation.
%at the leading order.
%and therefore is ignored.

%There are two cutoff wavenumbers for this case
%if the system is linearly unstable.
%They are 
The cutoff wavenumber for this case %the Rayleigh-Taylor unstable case
%is obtained similarly to the previous one
%from Eq.~(\ref{eq:dispersion_relation}) 
is given by
\begin{equation}
 k_c^2 = 
 \left\{
 \begin{array}{c@{\quad}c@{\quad}c}
  \displaystyle \left(\frac{k_0^2}{2} - |b|\right)
  \left[- 1 + \sqrt{1 + 2 \frac{-|a| + |b| k_0^2}
      {(k_0^2 / 2 - |b|)^2}}\ \right] & 
  \mbox{for} & \displaystyle \frac{k_0^2}{2} > |b|, \\[5mm]
  \displaystyle \left(|b| - \frac{k_0^2}{2}\right)
  \left[1 \pm \sqrt{1 + 2 \frac{-|a| + |b| k_0^2}
      {(|b| - k_0^2 / 2)^2}}\ \right] & \mbox{for} & 
  \displaystyle \frac{k_0^2}{2} < |b|.
 \end{array}
 \right.
\label{eq:cutoff_RT}
\end{equation}
The second line of this equation suggests
the existence of two cutoff wavenumbers when $-|a| + |b| k_0^2 < 0$.
Recall that $-|a| + |b| k_0^2 < 0$ is the condition that
%corresponding to positive or 
the growth rate is negative at infinitesimal wavenumber.
Therefore, the instability starts around $k = 0$ for the first case
of Eq.~(\ref{eq:cutoff_RT}), whereas
%In this case, 
finite modes between the two cutoff wavenumbers are unstable
for the second.
The finite critical wavenumber where the instability starts
for the second case is $k_{crit} = \sqrt{|b| - k_0^2 / 2}$.

In seeking the critical condition for the stability,
it is desirable to vary the liquid depth $d_l$ independently.
In the dispersion relation (\ref{eq:dispersion_relation}),
we have three dimensionless parameters depending on $d_l$,
$k_0^2$, $G$ and $S$. %except $d$ and $K$.
%For simplicity, 
%As was done before,
%Here, we fix the value of $d$ and
%assuming $\alpha = 1$.
%as was done before.
Since $S$ is proportional to $d_l$,
we choose $S$ as a control parameter
and the remaining two parameters are scaled by $S$
to obtain new parameters independent of $d_l$:
%\begin{eqnarray} 
% \lefteqn{k_{max}^2 = - \frac{2 k_0^2 - |b|}{3}} \\
% &+& \left[\frac{2 k_0^6 + 6 |b| k_0^4 
%      + 3 (2 |b|^2 - 9 |a|) k_0^2 + 2 |b|^3}{54} 
%      + \frac{1}{6} \sqrt{\frac{-|a| k_0^2 \{4 k_0^6 
%      + 12 |b| k_0^4 + 3 (4 |b|^2 - 9 |a|) k_0^2 
%      + 4 |b|^3\}}{3}}\right]^{1 / 3}
% \nonumber \\
% &+& \left[\frac{2 k_0^6 + 6 |b| k_0^4 
%      + 3 (2 |b|^2 - 9 |a|) k_0^2 + 2 |b|^3}{54} 
%      - \frac{1}{6} \sqrt{\frac{-|a| k_0^2 \{4 k_0^6 
%      + 12 |b| k_0^4 + 3 (4 |b|^2 - 9 |a|) k_0^2 
%      + 4 |b|^3\}}{3}}\right]^{1 / 3}
% \nonumber
%\end{eqnarray}
%\begin{eqnarray}
% \left\{
% \begin{array}{rcc}
% |G^*| k_0^{2*} S > 3 |A| & \mbox{for} & |G^*| S^4 < k_0^{2*} \\
% \displaystyle \frac{(|G^*| S^4 + k_0^{2*})^2}{S^3} > 12 |A| & \mbox{for} & 
%  |G^*| S^4 > k_0^{2*}
% \end{array}
% \right.
%\end{eqnarray}
%\begin{equation}
% F > |E|
%\end{equation}
\begin{equation}
 G^* = \frac{G}{S^3}, \qquad
  \Pi^* = S^2 \Pi, \qquad
  k_0^{2*} = S^2 k_0^2 = \frac{12 \eta (d + \lambda)}{\rho^2 d^4}
  \frac{K}{1 + K + \lambda / d} 
  \Pi^*.
\end{equation}
For simplicity, here we ignore the effect of $K$
%we do not consider the dependence of $K$ on $d_l$ here
by assuming the local thermodynamic equilibrium $K \rightarrow \infty$
or $\displaystyle \frac{K}{1 + K + \lambda / d} = 1$
in Eqs.~(\ref{eq:Ak_0^2})
because $K \gg 1$ for $\alpha = 1$ as mentioned above.
%$K = 1.2 \times 10^3 \alpha$
%for the water/vapor system of $d_l = 10^{-4}$m.
%In the following, we express 

\begin{figure}[htbp]
\scalebox{0.8}{
\includegraphics{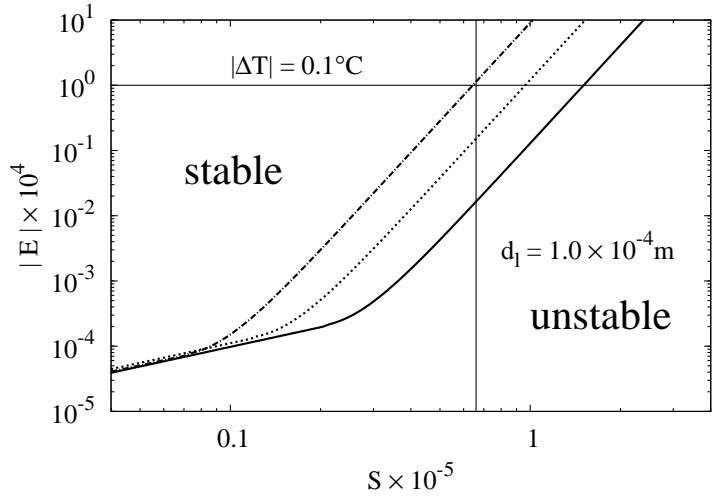}
}
\caption{\label{fig:stability_diagram}
Stability diagram in $S = \epsilon^{-3} \tilde{S}$ vs. 
$|E| = \epsilon^2 |\tilde{E}|$ plane 
for the Rayleigh-Taylor unstable case.
Here %$K \rightarrow \infty$, $\epsilon = 10^{-2}$ and 
$d = 0.5$, $1$ and $2$ from bottom to top.
The vertical line corresponds to $d_l = 1.0 \times 10^{-4}$m
and the horizontal one $|\Delta T| = 0.1\ {}^\circ$C.
}
\end{figure}

\begin{figure}[htbp]
\scalebox{0.8}{
\includegraphics{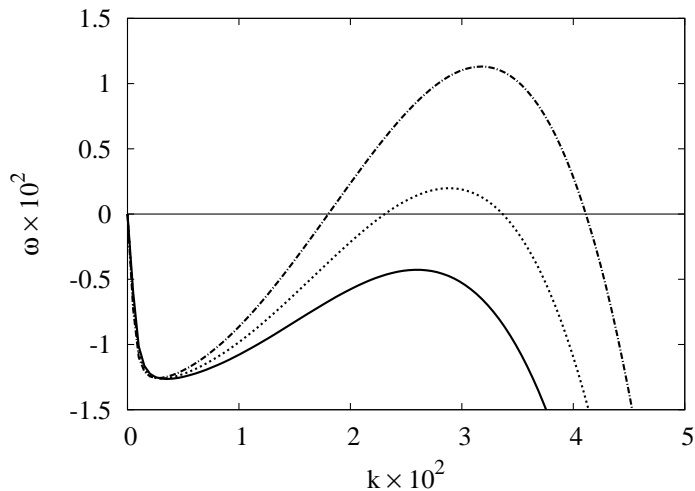}
}
\caption{\label{fig:dispersion_RT}
Growth rates $\omega$ versus wavenumber $k$
for the Rayleigh-Taylor unstable case.
Here $d = 2$, $|\Delta T| = 0.1\ {}^\circ \mathrm{C}$,
$|g| = 9.8\ \mathrm{m/s^2}$, $\alpha = 1$
%$K \rightarrow \infty$, $\epsilon = 10^{-2}$
and $d_l = 0.9 \times 10^{-4}$m, $1.0 \times 10^{-4}$m
and $1.1 \times 10^{-4}$m from bottom to top.
}
\end{figure}

The stability condition for the Rayleigh-Taylor instability
is obtained from the cutoff wavenumber (\ref{eq:cutoff_RT})
similarly to the previous case
and expressed in terms of the above new parameters as
\begin{eqnarray}
% F = 
 \left\{
 \begin{array}{cc@{\quad}c@{\quad}c}
 \displaystyle \frac{4 (d + \lambda)}{(4 + 3 d)(1 + d)}
 \frac{|G^*| \Pi^* S}{\rho |E|} & < 1 & 
  \mbox{for} & S < S_c, \\[5mm]
%\displaystyle \frac{|G^*| S^4}{k_0^{2*}} < 1, \\[5mm]
%  \displaystyle \frac{\rho^2 d^3 |G|}{12 \eta S \Pi} 
%  \frac{1 + K}{K} < 1, \\[5mm]
 \displaystyle \frac{\rho d^4}{\eta (4 + 3 d)(1 + d)} %\frac{1 + K}{K} 
  \frac{(|G^*| S^4 + k_0^{2*})^2}{12 |E| S^3} & < 1 & \mbox{for} & 
  S > S_c,
%  \displaystyle \frac{\rho^2 d^3 |G|}{12 \eta S \Pi} 
%  \frac{1 + K}{K} > 1.
%  \displaystyle \frac{|G^*| S^4}{k_0^{2*}} > 1.
%  |G^*| S^4 > k_0^{2*}.
 \end{array}
 \right.
\label{eq:RT_criterion}
\end{eqnarray}
where $S_c = (k_0^{2*}/|G^*|)^{1/4}$.
Here the two cases of this condition each correspond to those of
the cutoff wavenumber (\ref{eq:cutoff_RT}).
The first line represents the condition that $k_c^2$ is negative,
%The first line in this condition
which is essentially identical with 
the previous stability criterion (\ref{eq:criterion}),
while the second the one that $k_c^2$ is not real.
%the sum of the terms inside the root
%of Eq.~(\ref{eq:cutoff_RT}) is positive.
%itself does not have an imaginary part.
In Fig.~\ref{fig:stability_diagram} we plot the neutral stability curves
in S vs. \textbar E\textbar\ plane with the remaining parameters fixed
as shown in Table~\ref{tab:parameters_values}.
%Again, we set $K \rightarrow \infty$ and $\epsilon = 10^{-2}$.
%For $d_l = 1.0 \times 10^{-4}$m and $|\Delta T| = 0.01\ {}^\circ$C,
One can see the deflections of the curves,
which correspond to the transitional points ($S = S_c$)
between the two criteria in Eq.~(\ref{eq:RT_criterion}).
The stability curve for $d = 2$ passes near the point of 
$d_l = 1.0 \times 10^{-4}$m and $|\Delta T| = 0.1\ {}^\circ$C.
For this point, the system is stable for both $d = 0.5$
and $d = 1$.
%and unstable when $d = 2$.
From Fig.~\ref{fig:stability_diagram}
the stable region is wider for the thinner vapor layer,
suggesting the enhancement of
%the thinner vapor layer strengthens 
the stabilizing effect of lateral vapor pressure fluctuation.
Figure~\ref{fig:dispersion_RT} displays the dispersion curves
for the Rayleigh-Taylor unstable case
with $d = 2$ and $|\Delta T| = 0.1\ {}^\circ \mathrm{C}$,
$|g| = 9.8\ \mathrm{m/s^2}$, $\alpha = 1$
around $d_l = 1.0 \times 10^{-4}$m.
%As was pointed out, the shape of the curves differs from that of 
%Ref.~\onlinecite{bestehorn06:_regul_rayleig_taylor}
%in that the growth rate vanishes in the long-wave limit.
%Instead it resembles that in Fig.~2 of
%Ref.~\onlinecite{sultan05:_evapor},
%although the expression for the growth rate is different.
The fastest growing mode can be obtained from Eq.~(\ref{eq:fastest})
with $a, b < 0$.

\begin{table}[htbp]
\caption{\label{tab:parameters_values}
Values of the dimensionless parameters for the water/vapor
 system at $100\ {}^\circ \mathrm{C}$ and 1 atm
under terrestrial conditions.
Here, the tildes are restored to avoid confusion.
}
\begin{center}
\begin{tabular}%{c@{\quad \qquad}c@{\quad \qquad}c@{\quad \qquad}c}
{l@{\hspace{1.5cm}}l@{\hspace{1.5cm}}l}
\hline \hline \\[-6mm]
% $\Delta T$ & $0.1\ {}^\circ \mathrm{C}$ &  
% $d_l$ & $10^{-5}$ m &
 $\tilde{\rho} = 6.3 \times 10^{-4} \epsilon^{-2}$ &
 $\tilde{\eta} = 4.5 \times 10^{-2} \epsilon^{-1}$ & \\[2mm]
% $\tilde{E}$ & $1.0 \times 10^{-4} \epsilon^{-2}$ & 
% $\tilde{S}$ & $6.6 \times 10^3 \epsilon^3$ & 
 $|\tilde{G}^*| = 3.8 \times 10^{-13} \epsilon^{-8}$ &
 $\tilde{\Pi}^* = 6.5 \times 10^{-3} \epsilon$ &
 $\tilde{k}_0^{2*} = 8.7 \times 10^3 \epsilon^4 \displaystyle
 \frac{d + \lambda}{d^4} \frac{K}{1 + K + \lambda / d}$
 \\[5mm] \hline \hline
\end{tabular}
%\begin{tabular}{l@{\hspace{5mm}}l@{\hspace{5mm}}l@{\hspace{5mm}}l@{\hspace{5mm}}l}
%\hline \hline \\
%% $\Delta T$ & $0.1\ {}^\circ \mathrm{C}$ &  
%% $d_l$ & $10^{-5}$ m &
% $\tilde{\rho} = 6.3 \times 10^{-4} \epsilon^{-2}$ &
% $\tilde{\eta} = 4.5 \times 10^{-2} \epsilon^{-1}$ & %\\[5mm]
%% $\tilde{E}$ & $1.0 \times 10^{-4} \epsilon^{-2}$ & 
%% $\tilde{S}$ & $6.6 \times 10^3 \epsilon^3$ & 
% $|\tilde{G}^*| = 3.8 \times 10^{-13} \epsilon^{-8}$ &
% $\tilde{\Pi}^* = 6.5 \times 10^{-3} \epsilon$ &
% $\tilde{k}_0^{2*} = 8.7 \times 10^3 \epsilon^4 \displaystyle
% \frac{1}{d^3} \frac{K}{1 + K}$
% \\[5mm] \hline \hline
%\end{tabular}
\end{center}
\end{table}

\subsection{Effect of degree of nonequilibrium on the Rayleigh-Taylor instability}

\begin{figure}[htbp]
\scalebox{0.8}{
\includegraphics{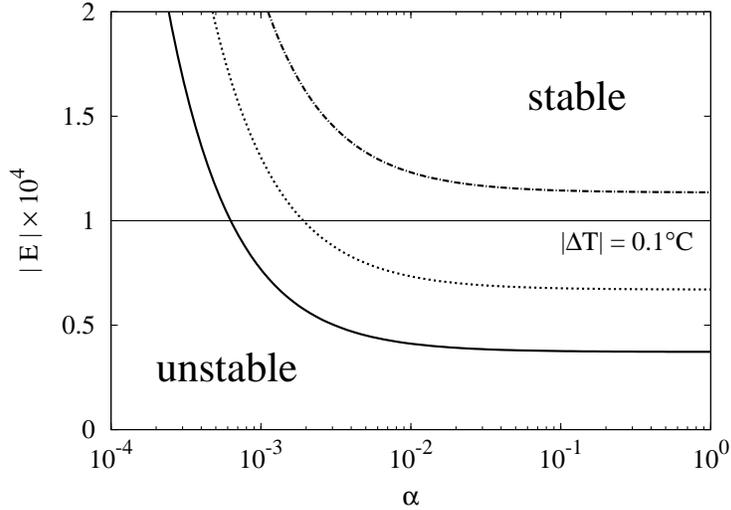}
}
\caption{\label{fig:accommodation}
Dependence of the stability on the accommodation coefficient.
Here $d = 2$ %, $\epsilon = 10^{-2}$
and $d_l = 0.8 \times 10^{-4}$m, $0.9 \times 10^{-4}$m and
$1.0 \times 10^{-4}$m from bottom to top.
The horizontal line indicates $|\Delta T| = 0.1\ {}^\circ \mathrm{C}$.
}
\end{figure}

In Fig.~\ref{fig:stability_diagram}, %we assumed the local thermodynamic
%equilibrium by taking the limit $K \rightarrow \infty$.
we did not consider the effect of nonequilibrium
because $K \gg 1$ for $\alpha = 1$ in the system considered.
However, the accommodation coefficient $\alpha$
can be much less than unity 
and thereby $K$ might approach one. 
%depending on the value of $\alpha$.
Here, we examine its effect %of the deviation from
%the local thermodynamic equilibrium 
on the stability of the system.
As was mentioned above, the degree of nonequilibrium $K$
does not enter the stability condition (\ref{eq:criterion})
for the thermodynamic unstable case.
For the Rayleigh-Taylor unstable case, however,
the criterion (\ref{eq:RT_criterion}) includes $K$
only in the lower line $(S > S_c)$ when $K$ is finite.
Then, it becomes
\begin{eqnarray}
% \begin{array}{c@{\quad}c@{\quad}c}
 \displaystyle \frac{\rho d^4}{\eta (4 + 3 d)(1 + d)}
 \frac{1 + K + \lambda / d}{K} 
  \frac{(|G^*| S^4 + k_0^{2*})^2}{12 |E| S^3} < 1 \quad \mbox{for} \quad 
  S > S_c.
%  |G^*| S^4 > k_0^{2*}.
% \end{array}
\label{eq:nonequilibrium}
\end{eqnarray}
Therefore, the coefficient of the left hand side of this inequality
increases as $K$ decreases, which may render the system unstable
even if it is stable at large $K$.
%from the stable state.
%The value of $K$ depends on the unknown accommodation coefficient $\alpha$.
In Fig.~\ref{fig:accommodation} we show the stability diagram
in $\alpha$ vs. \textbar E\textbar\ plane for the water/vapor
system. %, where $K = 1.2 \times 10^3 \alpha$ for $d_l = 10^{-4}$m.
For this system, the boundaries of the stability are almost constant
if $\alpha > 10^{-2}$.
The steep changes of the neutral stability curves are found for $\alpha <
10^{-2}$, where the system becomes unstable for the same value of
the evaporation number $E$.
The physical meaning of this behavior is that 
as the resistance to evaporation or condensation is increased
the stabilizing vapor pressure effect no longer overcomes
the destabilizing gravitational effect.
%Note that the threshold value of $\alpha$ for the change of 
%the neutral stability curve increases as the liquid depth $d_l$ decreases
%since $K$ is proportional to $\alpha d_l$.

\section{\label{sec:conclusion}Conclusion}

We have discussed the instability of thin liquid-vapor layers 
bounded by rigid parallel walls from both below and above.
In this system, the interfacial instability is induced by
lateral vapor pressure fluctuation, 
which is in turn attributed to the effect of phase change:
the vapor pressure becomes higher
at an evaporating portion of the interface and vice versa.
%the liquid evaporates at the hotter portion of the interface 
%and the vapor condenses at the colder one.
%evaporation occurs at the hotter portion of the interface
%and condensation at the colder one.
%The high vapor pressure drives the liquid away and 
%the low one pulls it up.
%In this system, if the liquid side is heated or the vapor side
%is cooled, an interfacial instability due to phase change occurs,
%which is induced by lateral vapor pressure fluctuation:
%and thereby the vapor pressure becomes higher there.
%in the evaporating region and vice versa.
%%the higher vapor pressure drives the liquid away
%%and the lower one pulls it up.
%According to this lateral vapor pressure gradient, 
The liquid is driven away
from the higher pressure place and pulled up to the lower pressure one.
This pressure-induced instability mechanism has not been considered
in the past.
%in the studies of evaporating or condensing thin liquid films, 
%since the previous works assume 
%the uniform ambient vapor pressure.\cite{burelbach88:_nonlin,bestehorn06:_regul_rayleig_taylor}
%In contrast, the surface deflection due to local mass loss
%or gain of the liquid layer can be neglected
%because of the very small vapor density compared to
%the liquid one.

%In order to formulate this problem,
In the formulation,
%At the liquid-vapor interface, 
the interfacial boundary condition
taking into account the pressure dependence of 
the local saturation temperature, Eq.~(\ref{eq:clapeyron}), was imposed.
The relation (\ref{eq:clapeyron}) is an extension of the kinetic law
and the condition of local thermodynamic equilibrium,
conventionally used in the literature.
We applied the long-wave approximation to both liquid and vapor
layers, assuming that the layers are too thin for thermal convection
to occur.
%ignoring convection.
The choice of scalings we adopted (\ref{eq:scalings}) allows us to decouple
the mass flux balance equation (\ref{eq:mass_flux}),
resulting in the neglect of mass loss or gain through the interface
by evaporation or condensation at the leading order.
In the dimensional form, we can formally write the decoupled equations
(\ref{eq:liquid_mass_flux}) and (\ref{eq:vapor_mass_flux}) as
\begin{eqnarray}
 \left\{
  \begin{array}{l}
   0 = \rho_l (\mathbf{v}_l \cdot \mathbf{n} 
    - \mathbf{v}_I \cdot \mathbf{n}), \\[1.5mm]
   J = \rho_v \mathbf{v}_v \cdot \mathbf{n},
  \end{array}
 \right.
\label{eq:decoupled_equations}
\end{eqnarray}
where the first and second equations correspond to
the leading-order and next-order equations in $\epsilon$.
The condition for this approximation is $\rho \ll \eta \ll 1$:
$\rho \ll \eta$ for the neglect of the effect of mass loss or gain
and $\eta \ll 1$ for that of the interface velocity
in the second equation of Eq.~(\ref{eq:decoupled_equations}).
To our knowledge, this decoupling approximation of the mass flux balance
between two phases has never been considered
and might have a possibility to be applied to other phase-change
problems where $\rho \ll \eta \ll 1$ holds.

As a result, a set of the equations 
describing the temporal evolution of the interface 
of the liquid-vapor layers has been derived.
%One of the equations (\ref{eq:film_thickness_equation})
%is written in the conserved form, although our model involves
%a phase-change phenomenon.
%This is the peculiarity of our model
%since most models addressing two-phase interfaces with phase change
%have the effect of mass loss or gain.
%Meanwhile, the effect of the lateral vapor pressure gradient
%induced by phase change is included in Eq.~(\ref{eq:film_thickness_equation})
%and the vapor pressure is enslaved to the film thickness
%through Eq.~(\ref{eq:vapor_pressure_equation}).
One of the equations (\ref{eq:film_thickness_equation})
is written in the conserved form for the film thickness,
as is expected from the decoupling of the mass flux balance mentioned above.
Therefore, the total mass of the liquid layer
is conserved in our model.
On the other hand, the effect of the lateral vapor pressure gradient
induced by phase change is included in Eq.~(\ref{eq:film_thickness_equation})
and the vapor pressure is enslaved to the film thickness
through Eq.~(\ref{eq:vapor_pressure_equation}).
The result of the linear stability analysis of this model shows that
the presence of local saturation temperature variation
by the vapor pressure
%the pressure dependence of the local saturation temperature
mitigates the growth of long-wave disturbances.
%the saturation temperature is increased (decreased)
%in higher (lower) vapor pressure regions.
The instability is enhanced 
for the smaller initial thickness ratio $d$
and larger dynamic viscosity ratio $\eta$,
which increase the kinetic resistance to the vapor flow.
The role of the vapor flow is to mitigate the effect
of the lateral vapor pressure variation,
which is different from that described in Ref.~\onlinecite{ozen04}.
%for the thinner vapor layer. %, which suggests that
%the vapor pressure effect should have a critical role 
%in causing the instability.
We also determined the criterion for the linear stability
of the system and found that only slight temperature gradients are sufficient
to overcome the stabilizing gravitational effect
for the water/vapor system.

We also considered the Rayleigh-Taylor instability of the system.
%From the stability condition,
%the neutrally stable states turned out to be
%well within the experimentally feasible region.
Here, the stabilizing vapor pressure effect is balanced
with the destabilizing gravitational effect
under very small temperature difference between the plates.
Again, the thinner vapor layer strengthens 
the stabilizing effect of lateral vapor pressure fluctuation.
However, for this case %the vapor pressure effect might be 
%significantly weakened 
%the Rayleigh-Taylor instability might occur
the instability domain may be widened
if the accommodation coefficient $\alpha$ 
is below a certain critical value.
This value is about $10^{-2}$ for the water/vapor system
of $d_l = 10^{-4}$m.

In this paper, we addressed only the linear stability of our model.
The next step is to proceed to the nonlinear analysis of the model.
We shall investigate the behavior of the solution of the equations
in the nonlinear regime by means of numerical simulation.
Three dimensional computation of the model will %be also performed to 
reveal the possibility of occurrence of pattern formation
as reported in Ref.~\onlinecite{bestehorn06:_regul_rayleig_taylor}.

The validity of the long-wave approximation,
on which our model is based, should be examined.
To this aim, the comparison with the full numerical simulation 
is required.
In particular, we do not know the critical condition for the onset of
small-scale cellular convection in liquid-vapor layers, 
which is neglected within the framework of the long-wave approximation.
There should exist a critical temperature gradient or thicknesses of
the layers (critical Rayleigh number)
for the transition between the conductive and convective
states of the temperature fields
if the buoyancy effect is taken into consideration.
It would be also of interest to make a comparison with 
the existing full linear stability analyses\cite{ozen04,ozen06:_rayleig_taylor}
and even weakly nonlinear analysis\cite{ozen04:_weak} of bilayer systems.  
The effect of heat convection should be estimated
to clarify the role of the vapor flow, which is different
between theirs and ours.

Finally, we make two remarks on our model.
First, we set the scalings in $\epsilon$ 
on the dimensionless material parameters
as Eq.~(\ref{eq:scalings})
assuming their values for the water/vapor system 
at $100\ {}^\circ \mathrm{C}$ and 1 atm as a representative substance.
Hence, if the material properties are considerably changed
(e.g. near the critical point),
we must reset the scalings appropriate for the relevant values,
which may lead to different evolution equations.
Second, we can incorporate the effect of intermolecular forces
as disjoining pressure in the formulation.
This effect will be dominant for layers of thicknesses
below 100nm, as in Ref.~\onlinecite{joo00:_inter} and \onlinecite{lenz07:_compet}.
However, for this scale the application of continuum theory
to the vapor layer would not be valid,
because the mean free path of a gas molecule amounts to about 60 or 70nm
at atmospheric pressure and becomes much larger at reduced pressure.

\begin{acknowledgments}
 The author thanks Alexander Oron
%, Sadayoshi Toh, Tomoaki Kunugi and the participants of
% the 4th International Marangoni Association Conference (IMA4) held at
% Noda, Japan in 2008, for helpful discussions. 
 for numerous suggestions which improve the paper.
 The author's visit to his laboratory in
 Technion-Israel Institute of Technology was financially supported
 by the Bilateral International Exchange Program (BIEP)
 of the Global COE ``The Next Generation of Physics, Spun from Universality
 and Emergence'' from the Ministry
 of Education, Culture, Sports, Science and Technology (MEXT) of Japan.
 The author also is grateful to Sadayoshi Toh, Tomoaki Kunugi and 
 the participants of 
 the 4th International Marangoni Association Conference (IMA4) held at
 Noda, Japan in 2008, for helpful discussions.
\end{acknowledgments}

\appendix

\section*{Appendix: Effect of thermocapillarity}

In the main text, we ignore the thermocapillarity for simplicity.
Here we examine its effect on our model.
If the thermocapillarity is present,
we must add the thermocapillary term in the stress balance
at the interface (\ref{eq:stress}):
\begin{equation}
 J (\mathbf{v}_l - \mathbf{v}_v) + (p_l - p_v) \mathbf{n} 
  - (2 \eta_l \mathbf{E}_l - 2 \eta_v \mathbf{E}_v) \cdot
  \mathbf{n} + 2 \sigma H \mathbf{n} + (\nabla_s \sigma) \mathbf{t}
  = \mathbf{0},
\label{eq:Marangoni_stress}
\end{equation}
where $\nabla_s = \mathbf{t} \cdot \nabla$.
Accordingly, Eq.~(\ref{eq:tangential_stress}) is modified as
\begin{equation}
 \mathbf{t} \cdot (2 \eta_l \mathbf{E}_l - 2 \eta_v \mathbf{E}_v) \cdot
  \mathbf{n} = \nabla_s \sigma.
\end{equation}
Here we use the linear approximation for the dependence of 
the surface tension on the temperature,
\begin{equation}
 \sigma = \sigma_0 - \gamma (T_I - T_{sat}(p_0)),
\label{eq:temperature_dependence}
\end{equation}
where $\sigma_0$ is the surface tension
at the initial equilibrium temperature $T_{sat}(p_0)$,
corresponding to $\sigma$ in the main text.
The coefficient $\displaystyle \gamma = - \frac{d \sigma}{d T_I}$
is positive for most common substances.
Then, the nondimensionalized tangential stress balance equation
(\ref{eq:dimensionless_tangential_stress}) becomes
\begin{equation}
 \mathbf{t} \cdot (2 \mathbf{E}_l - 2 \eta \mathbf{E}_v) \cdot
  \mathbf{n} = - M \nabla_s T_I,
\label{eq:tangential_Marangoni}
\end{equation}
where the Marangoni number is defined by
\begin{equation}
 M = \frac{d_l \rho_l \gamma \Delta T}{\eta_l^2}.
\end{equation}
%If the Marangoni effect is to be included in the model,
In order to include the thermocapillary effect in the model,
the Marangoni number is scaled as
\begin{equation}
 M = \epsilon^{-1} \tilde{M},
\end{equation}
under the long-wave approximation.
Then, Eq.~(\ref{eq:leading_order_tangential_stress}) reduces to
\begin{equation}
 \partial_z u_l = \eta \partial_z u_v - M \partial_x T_I,
\label{eq:leading_order_Marangoni}
\end{equation}
where the tilde over $M$ is omitted.
%corresponding to Eq.~(\ref{eq:leading_order_tangential_stress}).
We note here that the temperature dependence of the surface tension
(\ref{eq:temperature_dependence})
also allows the tangential variation of the normal capillary stress
in Eq.~(\ref{eq:Marangoni_stress}).
However, this effect is safely neglected in the framework
of the long-wave approximation (see Ref.~\onlinecite{burelbach88:_nonlin}).
Using Eq.~(\ref{eq:leading_order_Marangoni}),
one of the coefficients in Eq.~(\ref{eq:coefficients}) is modified as
\begin{equation}
 c_1(x,t) = -\displaystyle \frac{1}{2} (1 + d - h) \partial_x p_v -
  h \partial_x p_l - M \partial_x T_I,
\end{equation}
while the other does not change.
As a result, %the Marangoni term appears
only one of the equations (\ref{eq:film_thickness_equation})
contains the Marangoni term:
\begin{equation}
 \partial_t h = \partial_x \left[\frac{h + 3 (1 + d)}{12} h^2 \partial_x p_v 
		 + \frac{h^3}{3} \partial_x (G h - S \partial_x^2 h)
		 + \frac{h^2}{2} M \partial_x T_I \right].
\label{eq:modified_equation}
\end{equation}
Here, the interfacial temperature gradient is calculated
from Eq.~(\ref{eq:interfacial_temperature}) as
\begin{eqnarray}
 \lefteqn{\partial_x T_I = \frac{1}{\displaystyle 1 + K h
  + \frac{\lambda h}{1 + d - h}} \left[-\frac{(1 + d) d \partial_x h}
  {(1 + d - h)^2} + \frac{K \Pi}
  {\rho E} \partial_x (h p_v)\right]}\hspace{14.5cm}
\label{eq:interfacial_temperature_gradient} \\
 {} - \frac{\partial_x h}
 {\displaystyle \left(1 + K h + \frac{\lambda h}{1 + d - h}\right)^2}
 \left[K + \frac{1 + d}{(1 + d - h)^2} \lambda\right]
  \left[-\frac{(1 + d)(h - 1)}{1 + d - h}
   + \frac{K \Pi}{\rho E} h p_v\right].
 \nonumber
\end{eqnarray}
Note that if we set $K = 0$,
Eq.~(\ref{eq:interfacial_temperature_gradient}) is
\begin{equation}
 \partial_x T_I = - \frac{(1 + d) (d + \lambda)}
  {(1 + d - h + \lambda h)^2} \partial_x h.
\end{equation}
%Note that
The Marangoni term with this temperature gradient
in Eq.~(\ref{eq:modified_equation})
%becomes
is identical with that
%with the vanishing conductivity ratio $\lambda = 0$
of the two-layer model without phase change
in Ref.~\onlinecite{vanhook97:_long_benar}.

Under Eqs.~(\ref{eq:modified_equation}) and
(\ref{eq:vapor_pressure_equation}),
the dispersion relation (\ref{eq:dispersion_relation})
is altered as
\begin{equation}
 \omega = \frac{A' k^2}{k^2 + k_0^2} - \frac{1}{3} k^2 (G + S k^2)
  + \frac{1 + d}{2 d} \frac{M k^2}{\displaystyle 1 + K + \frac{\lambda}{d}},
\label{eq:modified_dispersion_relation}
\end{equation}
where
\begin{equation}
 A' = \frac{\eta (1 + d)}{\rho d^4}
 \frac{K}{\displaystyle 1 + K + \frac{\lambda}{d}}
 \left[(4 + 3 d) E + \frac{6 M \Pi}{\rho}
   \frac{K}{\displaystyle 1 + K + \frac{\lambda}{d}}\right].
\label{eq:modified_A}
\end{equation}
Here only the terms in the first line
of Eq.~(\ref{eq:interfacial_temperature_gradient})
contribute to the Marangoni terms
in Eqs.~(\ref{eq:modified_dispersion_relation})
and (\ref{eq:modified_A}).
The Marangoni term in Eq.~(\ref{eq:modified_dispersion_relation})
arises from the surface deflection,
whereas that in Eq.~(\ref{eq:modified_A})
from the saturation temperature variation
due to the vapor pressure gradient.
If the Marangoni number $M$ vanishes, 
Eqs.~(\ref{eq:modified_dispersion_relation}) and (\ref{eq:modified_A})
are identical to Eqs.~(\ref{eq:dispersion_relation}) and (\ref{eq:A}).
We have Eq.~(\ref{eq:modified_dispersion_relation}) be the same form
as Eq.~(\ref{eq:dispersion_relation})
by introducing a new dimensionless parameter
\begin{equation}
 G' = G - \frac{3}{2} \frac{1 + d}{d}
 \frac{M}{\displaystyle 1 + K + \frac{\lambda}{d}}.
\label{eq:modified_G}
\end{equation}

We estimate the relative importance of the Marangoni effect
on $A'$ and $G'$ by numerically comparing the two terms
in Eqs.~(\ref{eq:modified_A}) and (\ref{eq:modified_G}).
For the water/vapor system 
at $100\ {}^\circ \mathrm{C}$ and 1 atm,
$\gamma = 2 \times 10^{-4}$ N/m${}^\circ$C,
so that $\tilde{M} = 23 \epsilon$ for $d_l = 10^{-4}$m 
and $\Delta T = 0.1\ {}^\circ$C.
Therefore, it follows that 
$\displaystyle \frac{6 M \Pi}{\rho (4 + 3 d) E} \frac{K}{1 + K + \lambda / d}
= 4.7 \times 10^{-4} \frac{K}{1 + K + \lambda}$ and
$\displaystyle \frac{3}{2} \frac{1 + d}{d} \frac{M}{G}
\frac{1}{1 + K + \lambda / d}
= 0.62 \frac{1}{1 + K + \lambda}$ for $d = 1$.
%From these values, 
If $K \gg 1$,
the thermocapillary effect seems to be negligible
in the linear regime, which agrees with the former results
on the two-phase problems.\cite{mcfadden07:_onset,onuki05:_dropl,ozen04,mcfadden09:_onset_of_oscil_convec_in}
Yet both values increase as $d_l$ decreases 
or $\Delta T$ increases because the former is proportional to $d_l^{-1}$
(independent of $\Delta T$) and the latter to $d_l^{-2}$ and $\Delta T$.
If $K$ decreases, the former decreases and the latter increases.
However, the latter does not diverge as does
the left hand side of Eq.~(\ref{eq:nonequilibrium}).
In both thermodynamic and Rayleigh-Taylor unstable cases,
where the vapor pressure and gravity effects counteract each other,
the Marangoni effect acts as amplifying the former
and diminishing the latter.
Finally, we note that we do not know whether the Marangoni effect is
negligible in the nonlinear regime,
which will be ascertained in the ongoing numerical analysis.

%\bibliography{thin_films}

\end{document}